%% file: ms.tex
\documentclass[12pt,preprint]{aastex}

\newcommand{\Msun}{${M_{\odot}}$}

\slugcomment{\today}

\shorttitle{X-ray Young Stars in Carina Nebula}
\shortauthors{Sanchawala et al.}

\begin{document}

\title{X-RAY EMITTING YOUNG STARS IN THE CARINA NEBULA}


\author{
        Kaushar Sanchawala\altaffilmark{}, 
        Wen-Ping Chen\altaffilmark{}, 
        Hsu-Tai Lee\altaffilmark{}}
\affil{National Central University, Taiwan}

\author{
        Yasuhi Nakajima\altaffilmark{}, 
        Motohide Tamura\altaffilmark{}}
\affil{National Astronomical Observatory of Japan}

\author{
        Daisuke Baba\altaffilmark{},
	Shuji Sato\altaffilmark{}}
\affil{Department of Astrophysics, Nagoya University, Japan}

\and

\author{
        You-Hua Chu\altaffilmark{}}
\affil{University of Illinois at Urbana Champaign, USA}



\begin{abstract}

We present a multiwavelength study of the central part of the Carina
Nebula, including Trumpler~16 and part of Trumpler~14.  Analysis of the 
{\it Chandra X-ray Observatory\/} archival data led to the identification 
of nearly 450 X-ray sources.  
These were then cross-identified with optical photometric and spectroscopic 
information available from literature, and with deep near-infrared ($JHK_s$) imaging
observations.  A total of 38 known OB stars are found to be X-ray emitters.  
All the O stars and early B stars show the nominal relation between the
X-ray and bolometric luminosities, $L_{\rm X} \sim 10^{-7} L_{\rm bol}$.  
A few mid- to late-type B stars are found to be associated with X-ray
emission, likely attributable to T~Tauri companions.  
We discovered 17 OB star candidates which suffer large
extinction in the optical wavebands.  Some 300 sources have X-ray and
infrared characteristics of late-type pre-main sequence stars.  Our sample
presents the most comprehensive census of the young stellar population in
the Carina Nebula and will be useful for the study of the star-formation 
history of this massive star-forming region.  We also report the finding of a
compact ($5\arcmin \times 4\arcmin$) group of 7 X-ray sources, all of
which highly reddened in near-infrared and most X-ray bright.  
The group is spatially coincident with the dark 'V' shaped dust lane bisecting the 
Carina Nebula, and may be part of an embedded association.  The distribution of 
the young stellar groups surrounding the \ion{H}{2} region associated with Trumpler~16 
is consistent with a triggering process of star formation by the 
collect-and-collapse scenario.  

\end{abstract}





\section{INTRODUCTION}

Massive stars have a profound influence on neighboring molecular clouds.
On the one hand, the powerful stellar radiation and wind from even a
single such star would sweep away nearby clouds and henceforth prevent
subsequent star formation.  On the other hand, the massive star may
provide ``just the touch'' to prompt the collapse of a molecular cloud
which otherwise may not contract spontaneously.  Whether massive stars
play a destructive or a promotive role in cluster formation conceivably
depends on the availability of cloud material within the range of action,
though the details have not been fully understood.  If massive stars by
and large suppress star formation, low-mass stars could exist in the
immediate surroundings only if the low-mass stars predated massive star
formation.  In both the Orion and Lacerta OB associations, \citet{lee05}
and \citet{leechen} have found an evidence of triggered star formation by
massive stars. The UV photons from massive stars appear to have ionized
adjacent molecular clouds and the implosive pressure then compresses the
clouds to form next-generation stars of various masses, often in groups,
with high star formation efficiencies. The process is self-sustaining and
an entire OB association may be formed as a result.

The Carina Nebula, also known as NGC\,3372, is a remarkable star-forming
region where the most massive stars known in the Milky-Way Galaxy
co-exist. The Nebula, which occupies about 4 square degree area on the
sky, contains at least a dozen known star clusters \citep{feinstein95}.
The clusters with photometric and spectroscopic data are : Bochum (Bo) 10
and 11, Trumpler (Tr) 14, 15 and 16, Collinder (Cr) 228, NGC\,3293 and 
NGC\,3324. Tr\,14 and Tr\,16 are the most populous and youngest star clusters and
located in the central region of the Nebula.  The distance modulus for Tr\,16, 
quoted from the literature, ranges from 11.8 \citep{levato} to 12.55
\citep[MJ93 hereafter]{mj93} and for Tr\,14, from 12.20 \citep{feinstein83}
to 12.99 \citep{morrell}. \citet{walborn} derived a distance of $2.5$ kpc
for Tr\,16 using $R =3.5$. \citet{crowther} derived a distance of $2.6$ kpc
for Tr\,14. \citet{walborn73} and \citet{morrell} concluded that the two
clusters are at slightly different distance, whereas \citet{turner} and
\citet{mj93} concluded the two clusters are at the same distance.  A distance 
of 2.5~kpc is adopted for our study.  All the
clusters listed above, contain a total of 64 known O-type stars, the
largest number for any region in the Milky Way \citep{feinstein95}. Tr\,14
and Tr\,16 include six exceedingly rare main-sequence O3 stars.  The
presence of these very young stars indicates that the two clusters are
extremely young.  The two clusters also contain two Wolf-Rayet stars which
are believed to be evolved from even more massive progenitors than the O3
stars (MJ93). Furthermore Tr\,16 is the parent cluster to the famous
luminous blue variable (LBV), $\eta$ Carinae, which is arguably the most
massive star of our Galaxy (MJ93). With such a plethora of unusually
massive stars, the Carina Nebula is a unique laboratory to study the
massive star formation process, and the interplay among massive stars,
interstellar media and low-mass star formation.

In recent years, X-ray surveys have been very successful in defining the
pre-main sequence population of young star clusters \citep{fei02}. X-ray
emission has been detected from deeply embedded class I Young Stellar
Objects (YSOs) to low-mass pre-main sequence (PMS) stars of T Tauri types,
and intermediate-mass pre-main sequence stars of Herbig Ae/Be types to the
zero-age main-sequence stars. For late type main-sequence stars,
starting from late A to K and M dwarfs, the X rays are produced in the
very high temperature gas in the corona, which is thought to be heated due
to the dynamo magnetic fields \citep{maggio}. Massive stars, of O and
early B types, on the other hand emit X rays which are produced in the
shocks due to hydrodynamic instabilities in their radiatively driven
strong stellar winds \citep{lucy82}. The X-ray emission from the
classical T Tauri stars (CTTSs) or weak-lined T Tauri stars
(WTTSs) is believed to be thermal emission from the gas rapidly heated
to temperatures of the order $10^7$~K due to magnetic reconnection events
similar to the solar magnetic flares, but elevated by a factor of $10^1$
--$10^4$ \citep{fei99}. A recent work by \citet{preibisch05} presents the 
correlation of the X-ray properties with different stellar parameters, for 
a nearly complete sample of late-type PMS stars in the Orion Nebula Cluster.  
They concluded that the origin of X-ray emission in
T Tauri stars seems to be either a turbulent dynamo working in stellar
convection zone, or a solar like $\alpha$-$\Omega$ dynamo at the base of
the convection zone if T Tauri stars are not fully convective.  Among the
existing methods to identify the young stellar populations in a young star
cluster, the use of X-ray emission, which is nearly independent of the
amount of circumstellar material around the young stars \citep{fei02}, is
the least biased, especially in selection of weak-lined T Tauri stars which
lack the standard signatures of pre-main sequence stars such as the infrared
excess or strong $H\alpha$ emission lines.

In this paper we used the {\it Chandra X-ray Observatory\/} archival data 
of the Carina Nebula to
select the young stellar populations of the region. We then made use of
the optical photometric and spectroscopic information available in the
literature to identify the counterparts of the X-ray sources. We found
that more than 2/3 of the X-ray sources do not have any optical
counterparts. To characterize these sources further, we used the
Simultaneous InfraRed Imager for Unbiased Survey (SIRIUS) camera, mounted
on the Infrared Survey Facility, South Africa to carry out $J$, $H$,
and $K_{s}$ band imaging observations. Figure~1\footnote{Figures with better
resolution can be obtained from http://cepheus.astro.ncu.edu.tw/kaushar.html} 
shows the 
optical image of the Nebula from Digitized Sky Survey ($\sim$  25\arcmin 
$\times$ 25\arcmin), with 
the $Chandra$ field marked by a square centered on Tr 16 and covering 
part of Tr 14, which is in the north west of Tr 16. The  
field observed in the near infrared is about the same as the field of the
optical DSS image. We discuss the X-ray and NIR
properties of the known OB stars of the region. We discovered 17 massive
star candidates on the basis of their NIR and X-ray properties  
similar to those of the known OB stars in the region.   
These candidate OB stars probably escaped previous detection because of 
their large extinction in the optical wavelengths.  Furthermore, 
we identified some 300 CTTSs and WTTSs candidates, again on the 
basis of their X-ray and NIR properties.  Our study therefore produces 
the most comprehensive young star sample in the Carina Nebula, 
which allows us to delineate the star formation history in this 
seemingly devastating environment.  
In particular we report the discovery of an embedded 
($A_{\rm V} \sim 15$~mag) young stellar group located to the 
south-east of Tr\,16, and sandwiched between two dense molecular clouds.  
Similar patterns of newly formed stars in between clouds seem to   
encompass the Carina Nebula, a manifestation of the triggered star formation 
by the collect-and-collapse process \citep{deh05}.   
      
The paper is organized as follows. \S 2 describes the $Chandra$
and the NIR observations and the data analysis.  In \S 3, we present 
the cross-identification of $Chandra$ sources with the optical spectroscopic 
information (available in the literature) and with our NIR sample.  We discuss 
the results and implication of our study in \S 4.  \S 5 summarizes our results.

\section{OBSERVATIONS AND DATA REDUCTION} 

\subsection{X-ray data --- $Chandra$ }

The Carina Nebula was observed by the ACIS$-$I detector of {\it Chandra
X-ray Observatory}.  There were two observations in 1999 September 6,
observation ID 50 and 1249 (Table~1).  We began our data analysis with the
Level 1 processed event and filtered cosmic-ray afterglows, hot pixels,
$ASCA$ grades (0, 2, 3, 4, 6) and status bits.  Charge transfer
inefficiency (CTI) and time-dependent gain corrections were not applied,
because the focal plane temperature of these two observations was not
-120~C.  Because of the background flaring at the beginning of observation
of obs ID 50, we used a reduced exposure time of 8.5~ks.  Therefore, the
total exposure time of the two combined observations is 18120~s.  The
filtering process was done using the $Chandra$ Interactive Analysis of
Observations (CIAO) package and following the Science Threads from
$Chandra$ X-Ray Center.  We also restricted the energy range from 0.4 to
6.0~keV.  This would optimize the detection of the PMS stars and reduce
spurious detections.  Finally, we merged two observations to one image
(Figure 2), which was used for source detection.

WAVDETECT program within CIAO was utilized to detect sources in the merged
image. We ran wavelet scales ranging from 1 to 16 pixels in steps of
$\sqrt{2}$ with a source significance threshold of 3$\times$10$^{-6}$.  
Removing some spurious detections, e.g. some sources around partial shell
of X-ray emission surrounding $\eta$ Carinae \citep{sew01} and along the
trailed line due to $\eta$ Carinae itself, we got 454 sources eventually.
By using the merged image for source detection, we detected more than 
double the number of sources than reported by \citet{evans}.

We extracted the count of each source from the circular region centered on 
the WAVDETECT source position with a 95\% encircled energy radii ($R$(95\%EE)) 
\citep{fei02}. For the background determination, an annulus around each source
between 1.2 and 1.5 $R$(95\%EE) was used. Before
extracting the source counts from each observation, exposure and
background maps were created. An exposure map was computed to take into 
account vignetting and chip gaps, and an energy range of 1.2~keV was used 
for generating the exposure map. To avoid any sources within the 
background annuli for a given source, a background map was created 
excluding the sources in $R$(95\%EE).  This background map was used to obtain  
the source counts. We utilized DMEXTRACT tool of CIAO to 
extract source counts for each of the two observations. The total count of  
each source was then computed by combining the two observations. Finally 
the count rates were calculated for a total exposure time of 18120~s.  
The typical background count across the $Chandra$ field had a 3-$\sigma$ 
error of $\sim 1$~count.  

\subsection{Near-Infrared Data --- SIRIUS}

We carried out near-infrared imaging observations toward the Carina Nebula 
using the SIRIUS (Simultaneous InfraRed Imager for Unbiased Survey) camera 
mounted on the Infrared Survey Facility (IRSF) 1.4 m telescope, 
in Sutherland, South Africa. The SIRIUS camera
\citep{nagayama} is equipped with three HAWAII arrays of $1024 \times
1024$ pixels and provides simultaneous observations in the three bands,
$J$(1.25 $\mathrm{\mu m} $), $H$(1.63 $\mathrm{\mu m}$) and $K_s$(2.14
$\mathrm{\mu m}$) using dichroic mirrors. It offers a field of view of
$7.\arcmin8 \times 7.\arcmin8$ with a plate scale of $0.\arcsec45$
$\mathrm{pixel^{-1}}$.  In April 2003 nine pointings ($3 \times 3$) were observed
covering effectively $22\arcmin \times 22\arcmin$ and including the
$Chandra$ field. The central coordinates of the observed fields are $R.A.
= 10^h45^m05^s$ and $Dec. = -59\arcdeg 38\arcmin 52\arcsec$.  For each
pointing, 30 dithered frames were observed, with an integration time of 30
s each, giving a total integration time of 900~s.  Two pointings 
(\#5 and \#6) of the April 2003 data which suffered weather fluctuations 
were re-observed in January 2005, for which 45
dithered frames were observed with an integration time of 20~s, yielding
a total integration time of 900~s for each pointing. The typical seeing
during our observations ranged from $1.\arcsec0$--$1.\arcsec4$ and the
airmass from 1.2 to 1.5.  The standard stars No.~9144 and 9146 from 
\citet{persson} were observed for photometric calibration.

We used the IRAF (NOAO's Image Reduction and Analysis Facility) package to
reduce the SIRIUS data. The standard procedures for image reduction,
including dark current subtraction, sky subtraction and flat field
correction were applied.  The images in each band were then average-combined 
for each pointing to achieve a higher signal-to-noise ratio. We performed
photometry on the reduced images using IRAF's DAOPHOT package
\citep{stetson}. Since the field is crowded, we performed PSF (point
spread function) photometry in order to avoid source confusion. To
construct the PSF for a given image, we chose about 15 bright stars, well
isolated from neighboring stars, located away from the nebulosity and not
on the edge of the image.  The ALLSTAR task of DAOPHOT was then used to
apply the average PSF of the 15 PSF stars to all the stars in the image,
from which the instrumental magnitude of each star was derived. The
instrumental magnitudes were then calibrated against the standard stars
observed on each night.

\section{X-RAY SOURCES AND STELLAR COUNTERPARTS}

The optical spectroscopy of the stars in Tr\,14 and Tr\,16 has been done by
several groups, eg., \citet{walborn73,walborn82,levato,fitzgerald}.  The
latest work by MJ93 lists the brightest and the bluest stars of the two
clusters. We have used this list (Table~4 in MJ93) to find the
counterparts of our X-ray sources. Within a $3\arcsec$ search radius, our
cross-identification resulted in 30 OB stars from MJ93 as counterparts of
our X-ray sources. Apart from MJ93, we also checked for any possible
counterparts using SIMBAD\footnote{http://simbad.u-strasbg.fr/sim-fid.pl}.
This resulted into another 8 OB stars of the region \citep{tapia}, 
the spectral types of which were determined by the photometric Q method
\citep{json}.  

We also used our NIR data to search for the counterparts of the X-ray sources.
Again with a $3\arcsec $ search radius, we found counterparts for 432
sources. Thus, more than 95\% of the X-ray sources have NIR counterparts.
For 51 cases out of 432 sources, the NIR photometric errors are larger
than 0.1 mag in one or more bands. Most of these large photometric error
cases are for stars located in pointing 5, which is the Tr\,16 region. The
NIR photometry in this pointing is affected because of a large number of
bright stars and nebulosity around $\eta$ Carinae. Since we are using the
NIR colors of the sources to delineate their young stellar nature, an
uncertainty larger than 0.1~mag cannot serve the purpose.
Hence, in our analysis we consider only those cases for which the
photometric uncertainties are smaller than 0.1~mag in all the three bands, which
leaves us with 384 sources. For our analysis, we have converted the NIR
photometry into California Institute of Technology (CIT) system using the
color transformations between the SIRIUS and CIT systems as given in
\citet{nakajima}.

\section{RESULTS AND DISCUSSION}

\subsection{Known OB stars}

Table~2 lists the X-ray sources cross-identified with known OB
stars.  The coordinates of each X-ray source are listed in columns (1) and
(2), followed by the identifier of the optical counterpart of the
X-ray source, listed in column (3).  The optical $B$, and $V$ magnitudes, 
and the spectral type, listed in columns (4)--(6), were  
adopted in most cases from MJ93 and in others from
\citet{tapia}.  The color excess of each source, $E(B-V)$, given in
column (7), was also taken from MJ93, in which photometry and spectroscopy were 
used to estimate the intrinsic stellar $(B-V)_0$ \citep{fitzgerald70}.  
The bolometric magnitude, $M_{\rm bol}$, in column (8), was   
taken from \citet{massey01}.  For a small number of cases, where the
spectral types were taken from \citep{tapia}, the color excesses as well as the
bolometric magnitudes were estimated using their spectral types.  
Columns (9)--(11) list the IRSF NIR $J$, $H$ and $K_s$ magnitudes of the counterpart.  
Column (12) lists the X-ray counts of the
sources derived by the DMEXTRACT tool of the CIAO software, as described in 
\S 3.  We used the 
WebPIMMS\footnote{http://http://heasarc.gsfc.nasa.gov/Tools/w3pimms.html}
to derive the unabsorbed X-ray flux of the sources. To convert an X-ray
count to the flux, the Raymond Smith Plasma model with temperature 
$\log~T=6.65$, equivalent to $0.384 {\rm~keV}$, was adopted.  For 
the extinction correction, the color excess, $E(B-V)$ 
of each source was used to estimate the neutral hydrogen column density,
$N_H$. The X-ray flux is given in column (13), and the X-ray
luminosity, computed by adopting a distance of 2.5~kpc, is in column (14).  
The last column (15) contains the logarithmic ratio of the X-ray luminosity 
to the stellar bolometric luminosity, where the latter was derived from the 
bolometric magnitude, i.e., column (8).

Figure~3 shows the distribution of X-ray luminosities of the known OB stars in
our field.  Most OB stars have $\log L_{\rm X} \ga 31{\rm ~ergs~s^{-1}}$ with 
the distribution peaking $\sim \log L_{\rm X} = 31.7 {\rm~ergs~s^{-1}}$.  
The Wolf-Rayet star (HD\,93162) is
the brightest X-ray source in the sample, with 
$\log L_{\rm X} = 34.12 {\rm ~ergs~s^{-1}}$.  This star has been known to be unusually
bright in X rays as compared to other W-R stars in the region \citep{evans}.  
Though it has been thought to be a single star, a recent W-R
catalog by \citet{hucht} lists it as a possible binary 
(see discussions in \citet{evans}).

Among the 38 X-ray OB stars, there are 3 B3-type, 3 B5-type and 1 B7-type stars.  
Mid- to late-B type stars are supposed to be X-ray quiet as they have neither strong enough 
stellar winds as in the case of O or early B stars, nor the convective zones to
power the chromospheric/coronal activities as in the case of late-type stars.  
However, mid- to late-B type stars have been found to be X-ray emitters in earlier
studies, e.g., \citet{cohen}.  The X-ray luminosities of the mid- to
late-type B stars in our sample are comparable to those of T Tauri
candidates in the same sample.  Although this seems to provide circumstantial 
evidence of CTTS or Herbig Ae/Be companions to account for the X-ray 
emission, it does not rule out the possibility of a so far unknown emission 
mechanism intrinsic to mid- to late-B stars.   

The X-ray luminosities of the OB stars are known to satisfy the relation
with the stellar bolometric luminosities, namely, $L_{\rm X} \propto
10^{-7} L_{\rm {bol}}$. All but a few stars in our sample satisfy this
relation (Fig.~4). Among the outliers, labeled on the figure
by their spectral types, only HD\,93162 (a W-R star), and Tr\,16$-$22 (an
O8.5V star) are early type stars and hence their high $L_{\rm X}/L_{\rm
{bol}}$ is unusual. Tr\,16$-$22 is among the brightest X-ray sources in
our sample, with $\log L_\mathrm{X} = 32.83$~ergs~s$^{-1}$. It is
brighter in X rays by a factor of 5--20 compared to other O8.5V stars
and even brighter than the two O3 stars in the sample.  
\citet{evans} present a list of known binaries among the massive stars
and discuss the X-ray luminosities against their single or binary
status.  A massive companion may enhance the X-ray production by colliding
winds.  No binary companion is known to exist for either HD\,93162 or Tr\,16$-$22
\citep{evans} to account for their high X-ray luminosities and high 
$L_{\rm X}/L_{\rm {bol}}$ ratios.  The rest
of the X-ray sources which do not satisfy the correlation are mid-B or
late-B type stars.  A study by \citet{berghofer} about the X-ray
properties of OB stars using the $ROSAT$ database showed that the
$L_{\rm X}/L_{\rm bol}$ relation extends to as early as the spectral 
type B1.5, and inferred this as a possibly different X-ray emission 
mechanism for the mid- or late-B type stars as compared to the O or 
early-B stars.  In our sample, there are 3 B3 type stars which 
seem to satisfy this relation and all the stars later than B3 deviate 
significantly from the mean
value of $L_{\rm X}/L_{\rm {bol}}$ ratio for O and early-B stars.
  
\subsection{Candidate OB stars}

There are 17 anonymous stars in our sample which have similar NIR and X-ray
properties as the known OB stars in the region.  These stars appear to be massive stars of O
or B types, but we could not find their spectral type information in the 
literature, e.g., MJ93 or SIMBAD.  These candidate OB stars, with their optical and NIR
magnitudes along with their X-ray counts and X-ray luminosities are listed in Table 3.  
To determine their X-ray fluxes from counts, we made use of WebPIIMS.  For extinction 
correction, we used an average $E(B-V) = 0.52$ based on the Table~4 of MJ93, as we did 
not have the spectral class information to determine their individual
color excesses.  Other parameters to obtain the X-ray fluxes from the X-ray counts remain the
same as for the known OB stars.  We found that the use of an average value of $E(B-V)$, 
rather than the individual $E(B-V)$ values, in case of the known OB stars would make a 
difference of a factor of two or less in the derived X-ray luminosities.  Likewise for the temperature, 
using a $\log T$ between 6.4 to 7.1 also would make a difference of a factor of two or less 
in the X-ray luminosities.  Hence the use of an average color excess for candidate OB stars 
should not affect much our results.

Figure~5 shows the NIR color-color diagram of the known OB stars and the candidate OB stars.  
The solid curve represents the dwarf and giant loci \citep{bb}, and the parallel dashed lines 
represent the reddening vectors, with $A_J/A_V = 0.282$, $A_H/A_V = 0.272$, and $A_K/A_V = 0.112$
\citep{rieke}.  The dotted line indicates the locus for dereddened classical T Tauri stars 
\citep{meyer}.  It can be seen that the candidate OB stars are either intermixed with or 
redder than the known OB stars.  Figure~6 shows the NIR
color-magnitude diagram of the known and candidate OB stars.  The solid line represents the 
unreddened main sequence \citep{koornneef} at 2.5~kpc.  Some candidate OB stars are very bright 
in NIR, with a few even brighter than $K_s = 8$ mag.  In contrast, the candidate OB stars are 
fainter and redder than the known OB stars in the optical wavelengths (Figure~7), indicative 
of the effect of dust extinction, while both samples show a comparable range in X-ray luminosities 
(compare Figure~8 with Figure~2).  Thus, it appears that these candidate OB stars have escaped  
earlier optical spectroscopic studies because of their large optical extinction.  
Addition of these massive stars expands substantially the known list of luminous stars and thus
contributes significantly to the stellar energy budget of the region.

\subsection{PMS candidates}

Figures~9 and 10 show the NIR color-color and color-magnitude diagrams of all
the 380 X-ray sources with NIR photometric errors less than 0.1~mag.  By
using the criteria given in \citet{meyer}, we find about 180 stars as
CTTS candidates.  Apart from the CTTS candidates, the NIR colors suggest
quite a many possible weak-lined T Tauri star (WTTS) candidates.  The
X-ray and NIR data together hence turn up a large population of low-mass
pre-main sequence candidates.  The T Tauri candidates in our sample
(CTTS plus WTTS) should be a fairly secure T Tauri population, given
their X-ray emission and their NIR color characteristics.  Although much
work has been done on the massive stellar content in Tr\,14 and Tr\,16,
a comprehensive sample of the T~Tauri population has not been obtained
so far. \citet{tapia03} presented $UBVRIJHK$ photometry of Tr\,14,
Tr\,16 and two other clusters in the region, Tr\,15 and Cr\,232, and
noticed some stars with NIR excess in Tr\,14 and Tr\,16.  They estimated
the ages of Tr\,14 and Tr\,16 to be 1--6 million years.  To our
knowledge, our sample represents the most comprehensive sample of the
young stellar population in Tr\,14 and Tr\,16. The distribution of X-ray 
luminosities of the CTTS candidates is shown in Figure~11. Comparison with Figure~2 shows
that the X-ray luminosities of the T Tauri candidates are on the average
lower and hence consistent with the notion that late-type stars have
weaker X-ray emission.

\citet{fei05} pointed out that the X-ray luminosity functions (XLFs) of
young stellar clusters show two remarkable characteristics.  First, the shapes of
the XLFs of different young stellar clusters are very similar to each
other after the tail of the high luminosity O stars $(\log L_{\rm X} >
31.5 ~{\rm ergs~s^{-1}})$ is omitted.  Secondly, the shape of this 
'universal' XLF in the 0.5--8.0~keV energy range resembles a lognormal distribution with 
the mean, $\log L_{\rm X} \approx 29.5 ~{\rm ergs~s^{-1}}$ and the standard
deviation $\sigma(\log L_{\rm X}) \approx 0.9$ (see Figure~2 in \citet{fei05}).   

The $Chandra$ observations we used in this work include only part of
Tr\,14.  For Tr\,16, we can make an estimate of the total stellar
population in reference to the XLF of the Orion Nebula Cluster (ONC)
derived from the $Chandra$ Orion Ultradeep Project \citep{getman05}. 
The limiting X-ray luminosity of our sample is $L_{\rm X} \sim 30.5~{\rm
ergs~s^{-1}}$. Excluding the high X-ray luminosity tail, i.e., $L_{\rm
X} > 31.5~{\rm ergs~s^{-1}}$, which includes about 30 known OB stars and
candidate OB stars described earlier, the slope of the Tr\,16 XLF is
consistent with that of the ONC in the X-ray luminosity range of our
sample.  This suggests that our sample represents about 20\% of the
X-ray members in the cluster.  We hence estimate that the total stellar
population of Tr\,16 should be $\sim$ 1000--1300.  Furthermore, the
X-ray luminosities are known to be correlated with stellar masses, as
found in the $ROSAT$ data \citep{fei93} and also in the $Chandra$
studies of the ONC \citep{flaccomio03,preibisch05}.  Comparing the XLF
of Tr\,16 with the ONC XLF versus stellar mass (Figure~5 in
\citet{fei05b}), we infer that our sample is about 60\% complete for the
stars with masses larger than 1 \Msun, and 40\% complete between
0.3--1~\Msun.  Our deep NIR data covering clusters Tr\,14, Tr\,16 and
Cr\,232 would probe even lower mass end of the stellar population.  The
analysis of the complete NIR results will be presented elsewhere.

\subsection{A compact embedded X-ray group}

We notice a group of 7 X-ray sources concentrated in a field of 5\arcmin
$\times$ 4\arcmin, located south-east of Tr\,16 and coincident with the
prominent dark 'V' shaped dust lane which bisects the Carina Nebula. 
Adopting a distance of 2.5~kpc, the physical size of this star group is 
about 4~pc.  Each of these 7 sources has an NIR counterpart, listed 
in Table~6 with their coordinates, $J$, $H$, and
$K_s$ magnitudes and X-ray counts.  We use the star identification number
in column~1 of Table~6 in further discussion.  
The NIR colors have been used to estimate the neutral hydrogen column density.  
Stars in this group are bright and suffer large amounts of reddening, as seen in the  
NIR color-color and color-magnitude diagrams (Figures~12 and 13). 
The brightest sources (stars 4, 6 and 7) in NIR, with $K_s \sim$~8.5--10.5~mag,  
are also X-ray bright, with $L_{\rm X} \sim 10^{32}$
$\mathrm{ergs~s^{-1}}$.  Star 4 is a known O4 star \citep{rgsmith}.  Our
NIR magnitudes for this star match with those reported by \citet{rgsmith}.
Apart from star 4, we could not find any optical photometric or
spectroscopic information in the literature for the others sources.  
The bright NIR and X-ray stars 4, 5, and 6 can be clearly seen in the optical 
Digitized Sky Survey (DSS) image (Figure~14), whereas the other sources which are highly 
extincted even in NIR are not visible at all.  Figure~15 shows the IRSF $K_s$
image with the sources marked. Stars 1 and 5 are peculiar because they are highly 
extincted ($A_V \sim$ 15--25 mag estimated from their NIR colors), yet both are 
X-ray bright with $L_{\rm X} \sim 10^{33}$ $\mathrm{ergs ~ s^{-1}}$. 
What could be the nature of these sources?  Their NIR fluxes and colors,  
along with their non-detection in the DSS image, 
imply that they could be reddened T Tauri or class~I objects.
But their X-ray luminosities are much higher than observed in typical T
Tauri stars ($ < 10^{32} \mathrm{ergs~s^{-1}}$). One possibility
is that they are heavily embedded massive stars.  The rest two sources of
the group, stars 2 and 3 are relatively faint in both NIR and X rays,
thus appear to be reddened T Tauri stars.

It is worth noting that the above mentioned group is spatially close,
$\sim 7\arcmin$, to the deeply embedded object, IRAS\,10430$-$5931.
With $\mathrm{~^{12}CO(2-1)}$ and $\mathrm{~^{13}CO(1-0)}$ observations, 
\citet{megeath} found this $IRAS$ source associated with a bright-rimmed 
globule with a mass of $\sim 67$~\Msun.  They also found sources with 
NIR excess around this IRAS object and provided the first indication of 
star-formation activity in
the Carina region.  More recently, a mid-infrared study by
\citet{nsmith} discovered several clumps along the edge of the dark
cloud east of $\eta$ Carinae, including the clump associated with 
IRAS\,10430$-$5931.  They noted that each of these clumps is a potential site of
triggered star-formation due to their location at the periphery of the
Nebula behind the ionization fronts.  We compared the spatial
distribution of this group with the $\mathrm{~^{12}CO(1-0)}$
observations by \citet{brooks} in Figure~16.  The group of young stars 
is 'sandwiched' between two cloud peaks.  It is not clear whether
the group is continuation of Tr\,16 but obstructed by the dark dust lane, or
is a separate OB group/association still embedded in the cloud. 


Figure~17 shows all the $Chandra$ X-ray sources overlaid with the 
$\mathrm{~^{12}CO(1-0)}$ image \citep{brooks}.  One sees immediately a general 
paucity of stars with respect to molecular clouds.  Tr\,16 is ``sandwiched" 
between the north-west and south-east cloud complexes.  All the X-ray sources
(i.e., young stars) associated with these clouds in turn are seen either 
intervening between clouds or situated near the cloud surfaces facing Tr\,16.     
The morphology of young stellar groups and molecular clouds peripheral to 
an \ion{H}{2} region (i.e., Tr\,16) fits well the description of the 
collect-and-collapse mechanism for massive star formation, first proposed by 
\citet{elmegreen77} 
and recently demonstrated observationally by \citet{deh05,zav06}.  The expanding 
ionization fronts from an \ion{H}{2} region compress the outer layer of a nearby cloud 
until the gas and dust accumulate to reach the critical density for gravitational 
collapse to form next-generation stars, which subsequently cast out their own cavities.  
This collect-collapse-clear process may continue as long as massive stars are 
produced in the sequence and there is sufficient cloud material in the vicinity.

\section{SUMMARY}

We detected 454 X-ray sources in the $Chandra$ image of the Carina
Nebula observed in September 1999.  About 1/3 of the X-ray sources have
optical counterparts in the literature, including 38 known OB stars in 
the region.  In comparison our NIR observations detect counterparts for 
more than 95\% of the X-ray sources.  The X-ray luminosities of the known 
OB stars range in $\sim 10^{31}$--$10^{34}~{\rm ergs~s^{-1}}$, with the 
Wolf-Rayet star, HD 93162, being the strongest X-ray source with 
$\log L_{\rm X} = 34.12 \mathrm{~ergs~s^{-1}}$.   The W-R star also has a 
very high $L_{\rm X}/L_{\rm bol}$ ratio, $\sim -5.39$.  The only other 
early-type star with a high $L_{\rm X}/L_{\rm bol}$ ratio is an O8.5V 
type star, Tr\,16$-$22, which also has a very high X-ray luminosity 
of $L_{\rm X} = 32.83$ for its spectral type.  All other O and 
early B (up to B3 type) stars satisfy the canonical relation, 
$L_{\rm X} \sim ~10^{-7}~L_{\rm {bol}}$.  There are several 
mid- to late-B type stars emitting X rays with X-ray luminosities comparable 
with those typical for T~Tauri stars.  Hence, it is possible that the 
X-ray emission from these mid- and late-B stars is coming from T~Tauri 
companions.  We discovered 17 candidate OB stars which have escaped 
detection in previous optical studies because of the larger dust extinction 
they suffer.  These candidate OB stars have the same characteristics as known 
OB stars in terms of X-ray luminosities and NIR fluxes and colors.  
If most of them turn out to
be bona fide OB stars, this will be already half the number of the known
OB stars found as X-ray emitters in the region and would add
significantly to the stellar energy budget of the region.  The NIR colors
of the X-ray counterparts show a large population of low-mass pre-main
sequence stars of the classical T~Tauri type or the weak-lined T~Tauri type. 
Some 180 classical T~Tauri candidates are identified, whose X-ray 
luminosities range between $10^{30}$ to $10^{32}~{\rm ergs~s^{-1}}$, 
lower than those for OB stars.  Comparison of the X-ray luminosity 
function of Tr\,16---which is about 60\% complete for stars with masses 
1--3~\Msun and 40\% complete for 0.3--1~\Msun---with that of typical young star 
clusters suggests a total stellar population $\sim$ 1000--1300 in Tr\,16. 
A compact group of highly reddened, X-ray bright and NIR bright sources is found to 
the south-east of Tr\,16.  The group is associated with an $IRAS$ source and
coincident with the dust lane where many mid-IR sources have been
predicted to be the potential sites of triggered star-formation.  This star 
group is ``sandwiched" between two peaks of the $\mathrm{~^{12}CO(1-0)}$ emission.
Such star-cloud morphology is also seen in the peripheries of the \ion{H}{2} 
complex in Tr\,16, a manifestation of the collect-and-collapse triggering 
process to account for the formation of massive stars.

\acknowledgements
This publication makes use of the $Chandra$ observations of
the Carina Nebula made in September 1999. We made use of the SIMBAD Astronomical
Database to search the optical counterparts for the $Chandra$ X-ray sources.
We thank Kate Brooks for providing us with the $\mathrm{~^{12}CO(1-0)}$ data of the
Carina which was obtained with the Mopra Antenna, operated by the Australia
Telescope National Facility, CSIRO during 1996-1997. KS, WPC and HTL acknowledge
the financial support of the grant NSC94-2112-M-008-017 of the National Science Council of
Taiwan.

\clearpage

\input{tab1.tex}

\clearpage

\input{tab2.tex}

\clearpage

\input{tab3.tex}

\clearpage

\input{tab4.tex}

\begin{figure}
\begin{center}
\plotone{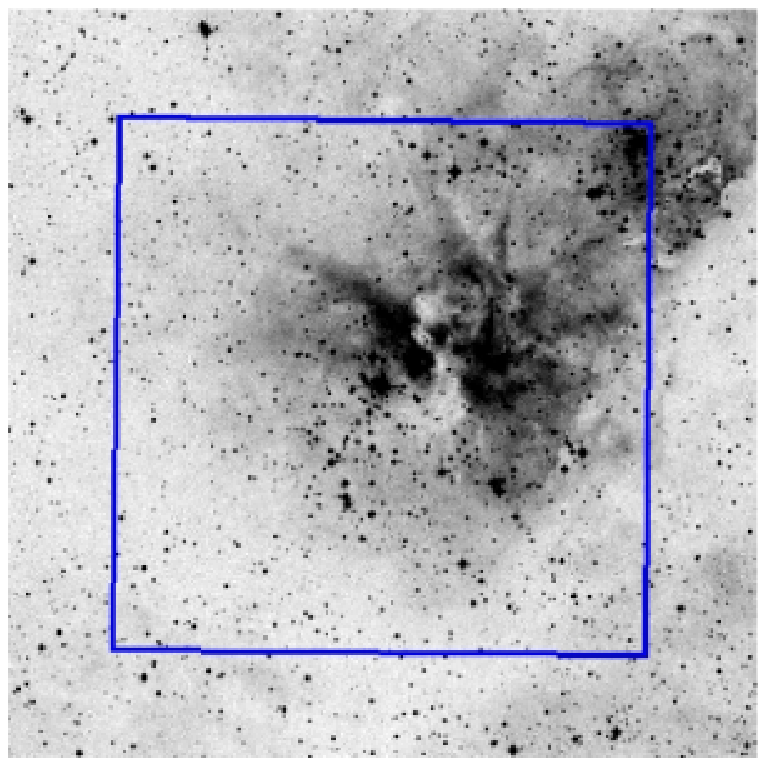}
\caption{The $Chandra$ field is marked by a square, centered on Tr 16 and
covering part of Tr 14, seen in the north west, in the DSS optical image
(25 \arcmin $\times$ 25 \arcmin). The field observed in NIR is about the same as
the DSS field shown here.
}

\end{center}
\end{figure}

\begin{figure}
\begin{center}
\plotone{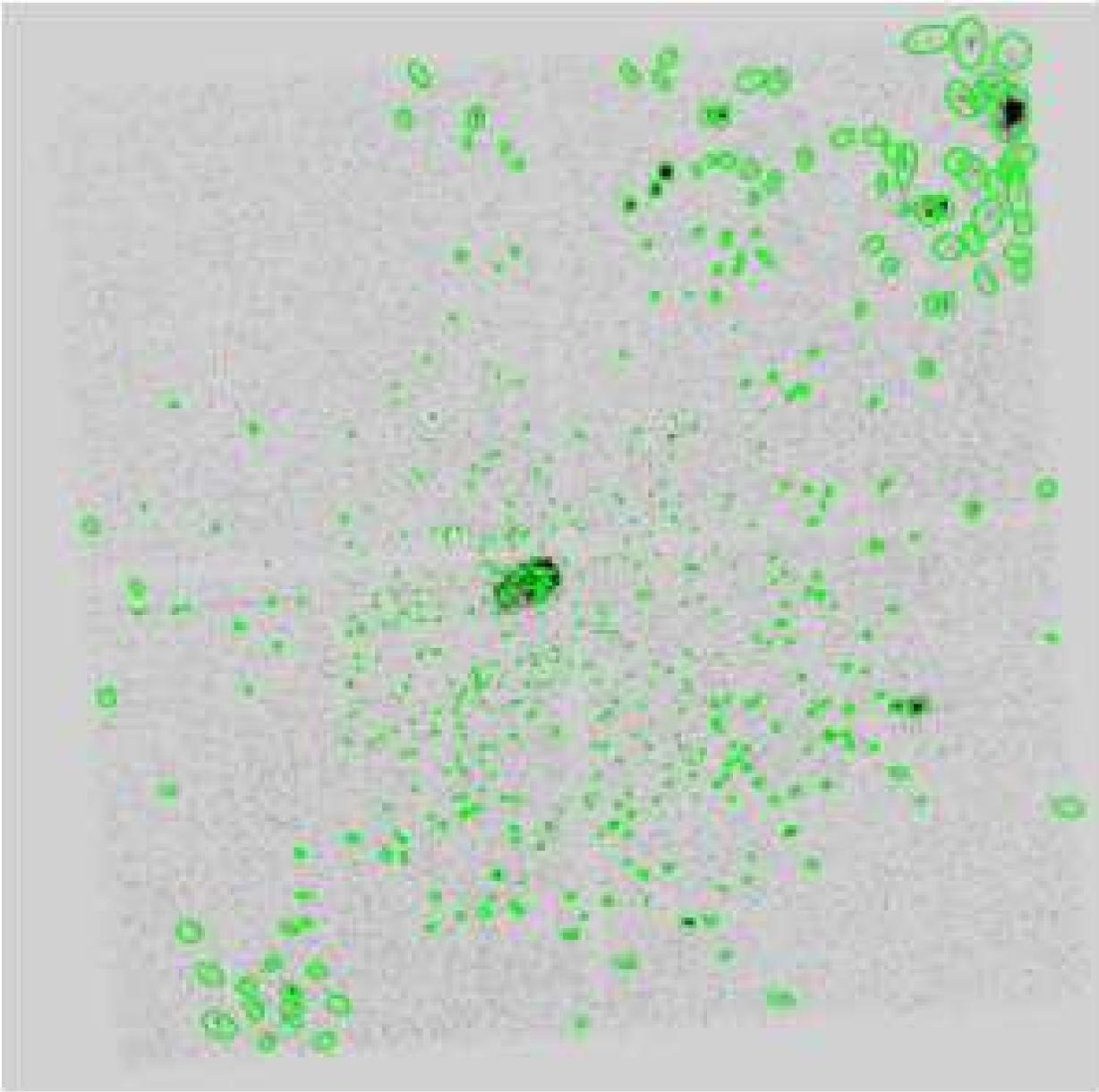}
\label{chandra}
\caption{
 The $Chandra$ ACIS-I image of the Carina Nebula, centered on $\eta$ Carinae. 
 The field of view of the image is 17\arcmin $\times$ 17\arcmin. The X-ray 
 sources are marked in ellipses, with the size of each ellipse being proportional 
 to the positional uncertainty of the source.}
\end{center}
\end{figure}

\clearpage

\begin{figure}
\begin{center}
\plotone{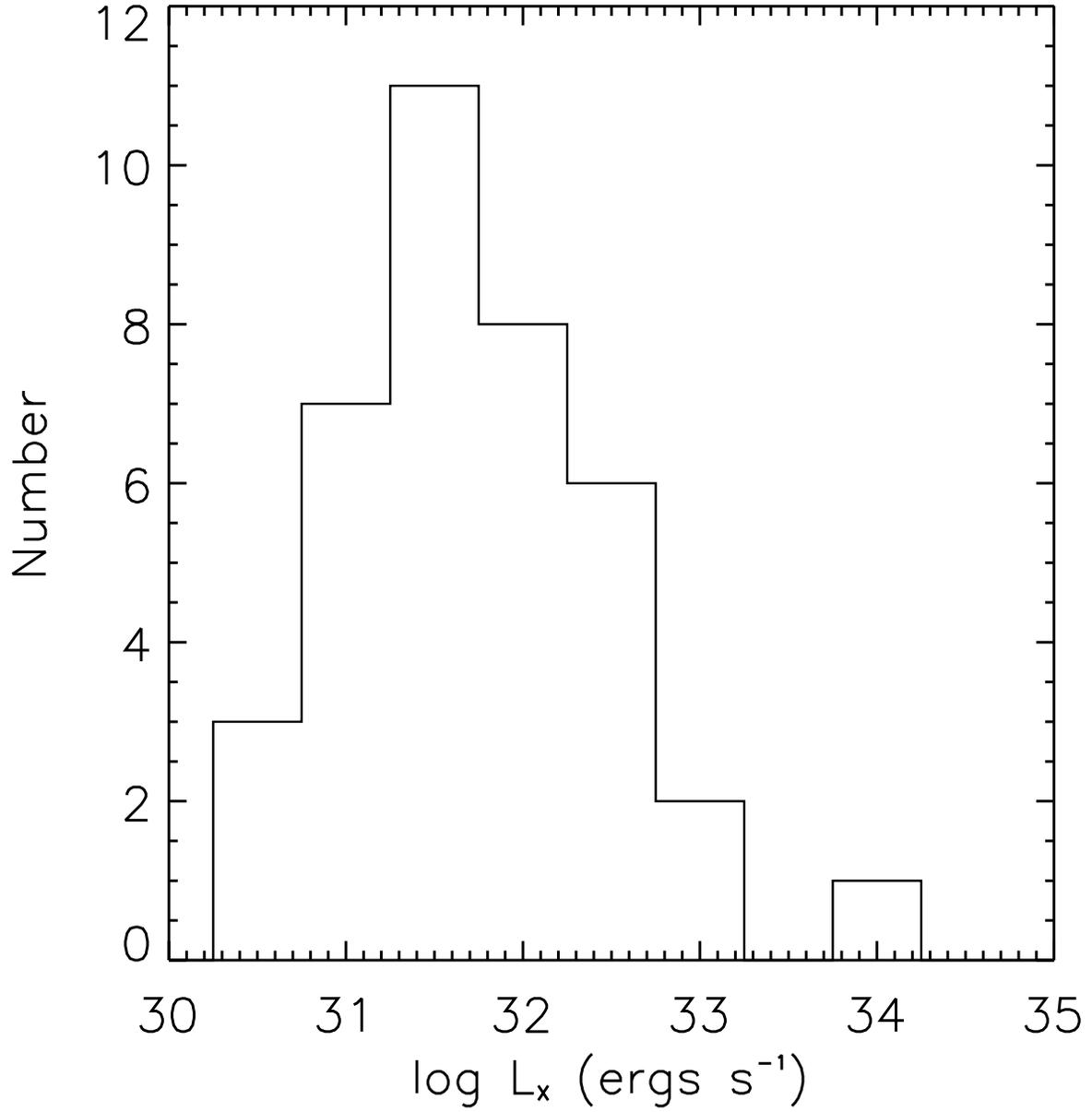}
\caption{
 The distribution of X-ray luminosities of known OB stars.}
\end{center}
\end{figure}

\clearpage

\begin{figure}
\begin{center}

\plotone{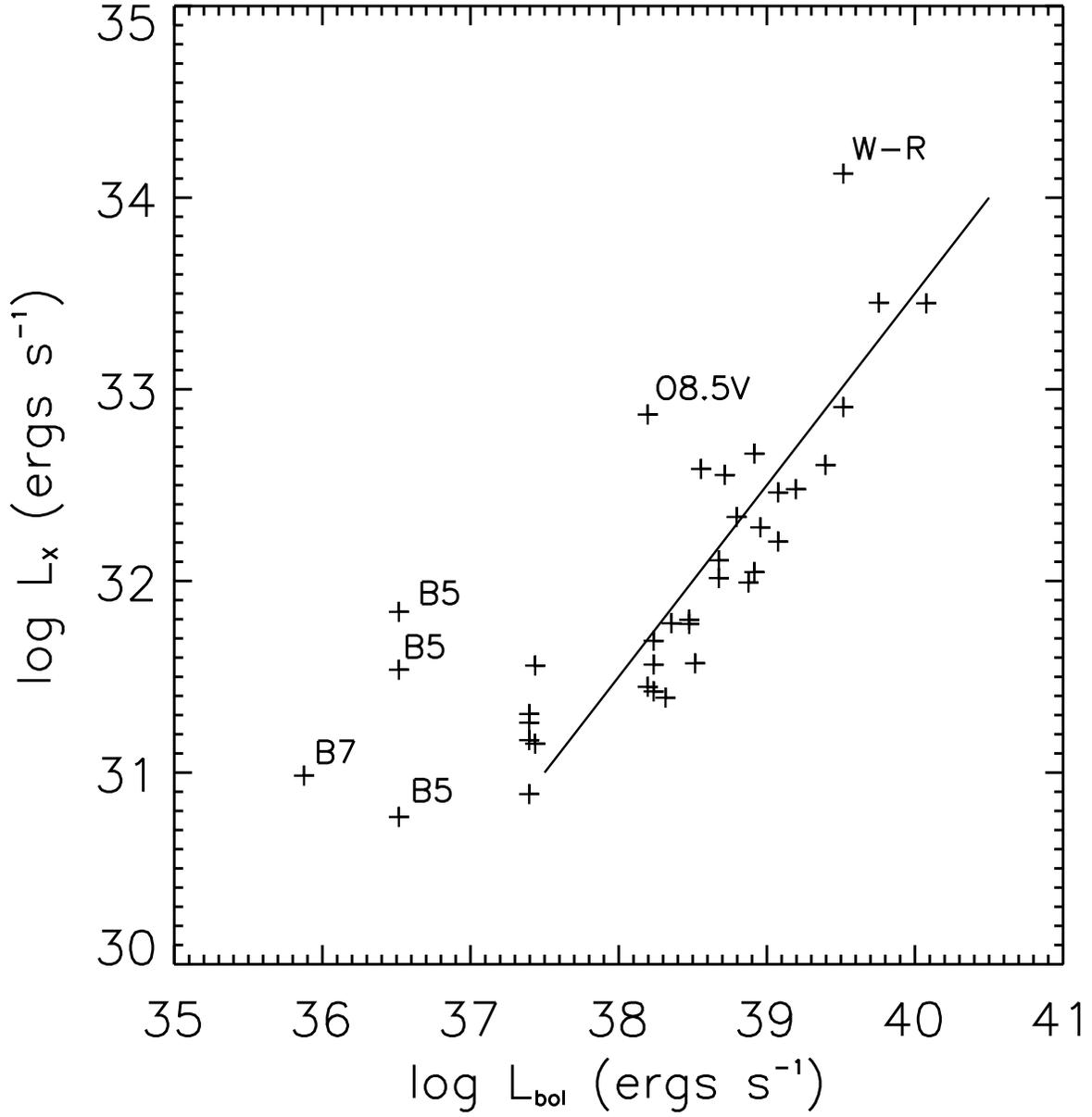}
\caption{
 The X-ray luminosity versus bolometric luminosity.  
 Most OB stars satisfy the well-known relation $L_{\rm X} \propto 
 10^{-7} L_{\rm bol}$.  The line is the linear regression fit with a 
 slope $\log L_\mathrm{X} = -6.5\, \log L_\mathrm{{bol}}$. 
 The outliers are labeled with their spectral types.}
\end{center}
\end{figure}

\clearpage

\begin{figure}
\begin{center}
\plotone{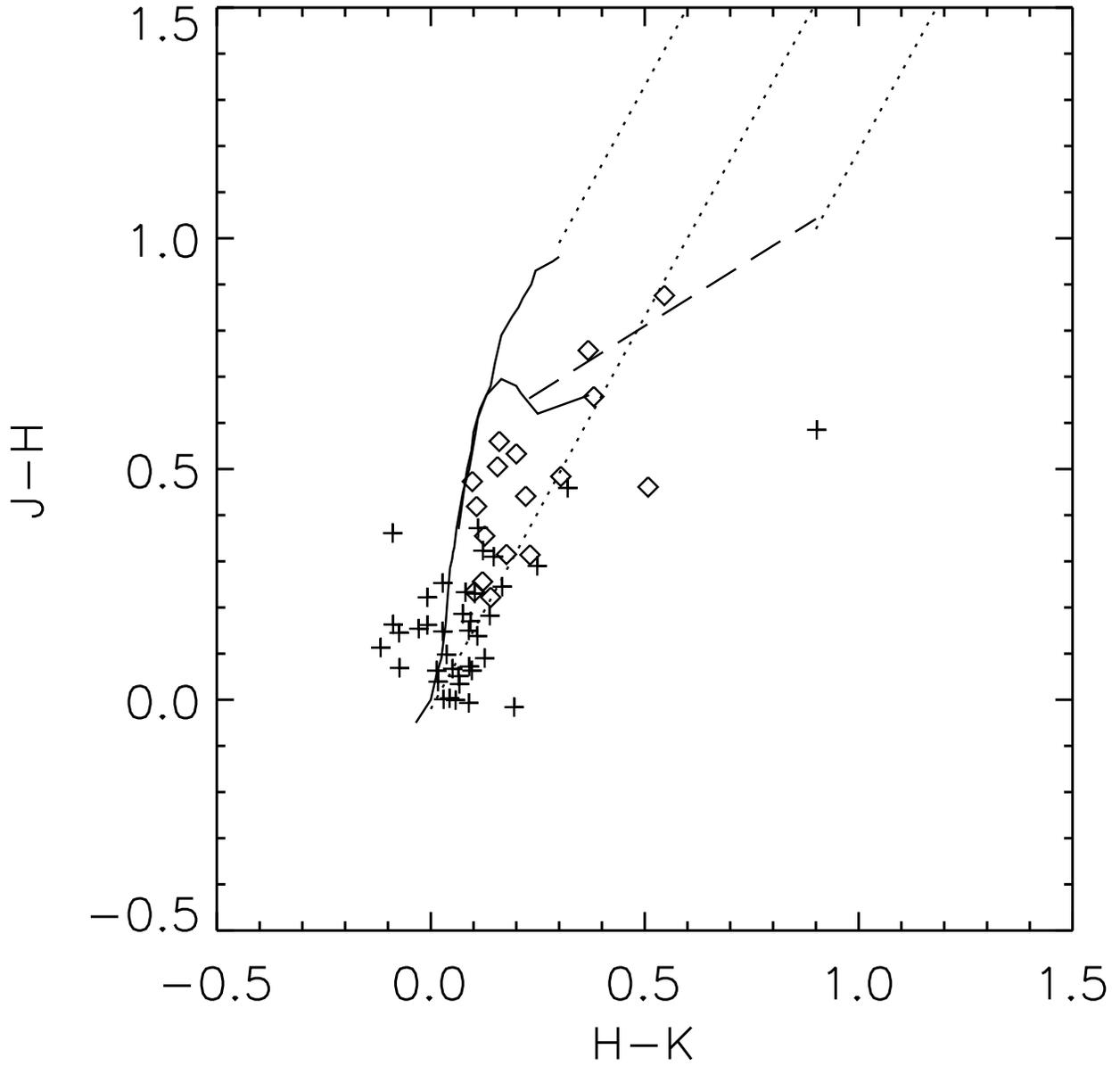}
\caption{
 The NIR color-color diagram of known OB stars (pluses) and 
candidate OB stars (diamonds). 
The dwarf and giant loci are shown as solid curves \citep{bb} and the
classical T Tauri locus is shown as the dashed line \citep{meyer}. 
The dotted lines represent the reddening band.}
\end{center}
\end{figure}

\clearpage

\begin{figure}
\begin{center}
\plotone{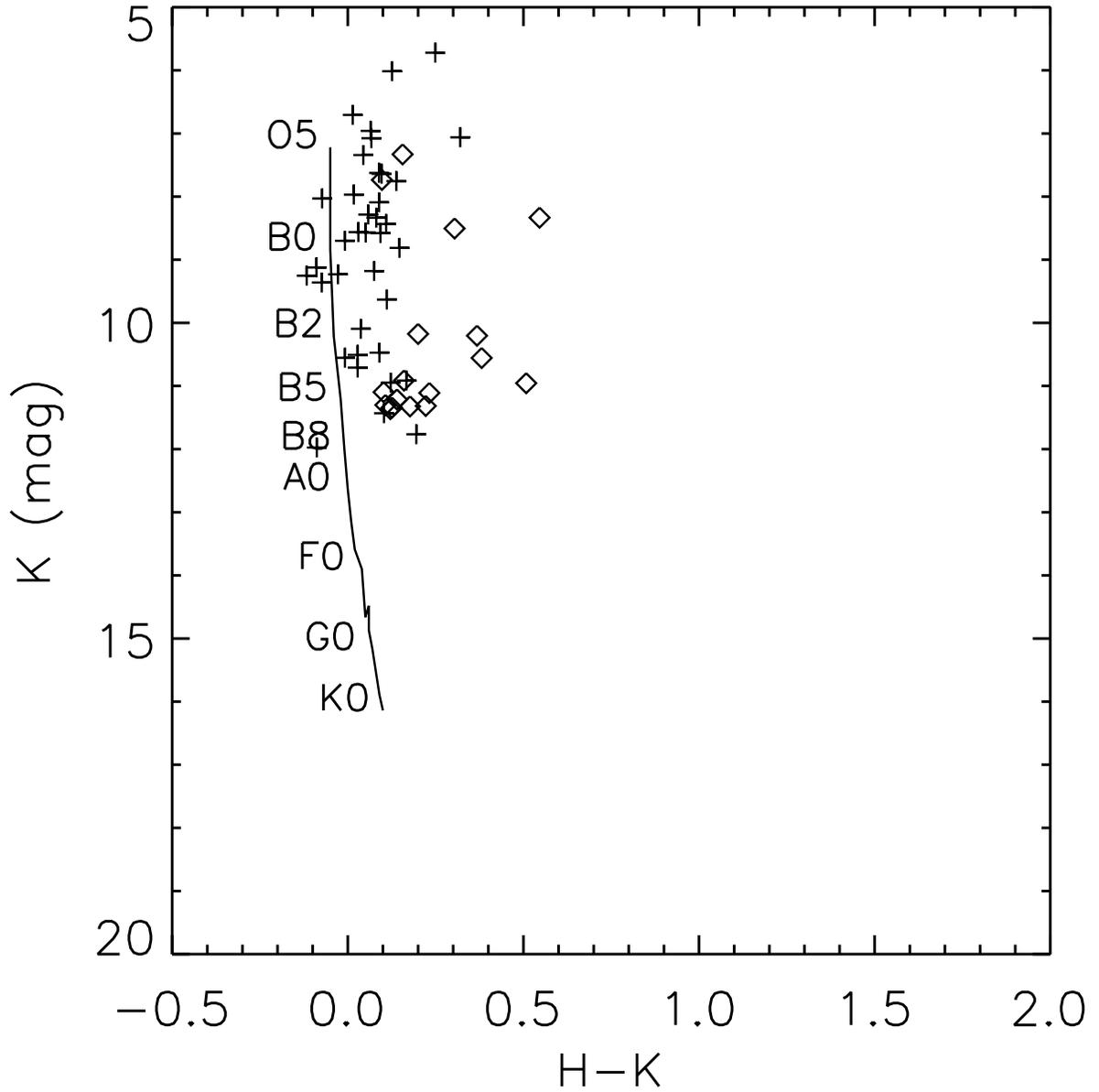}
\caption{The NIR color-magnitude diagram of known OB stars 
(pluses) and of candidate OB stars (diamonds). 
The vertical line represents the unreddened main sequence \citep{koornneef} at 
2.5~kpc.}
\end{center}
\end{figure}

\begin{figure}
\begin{center}
\plotone{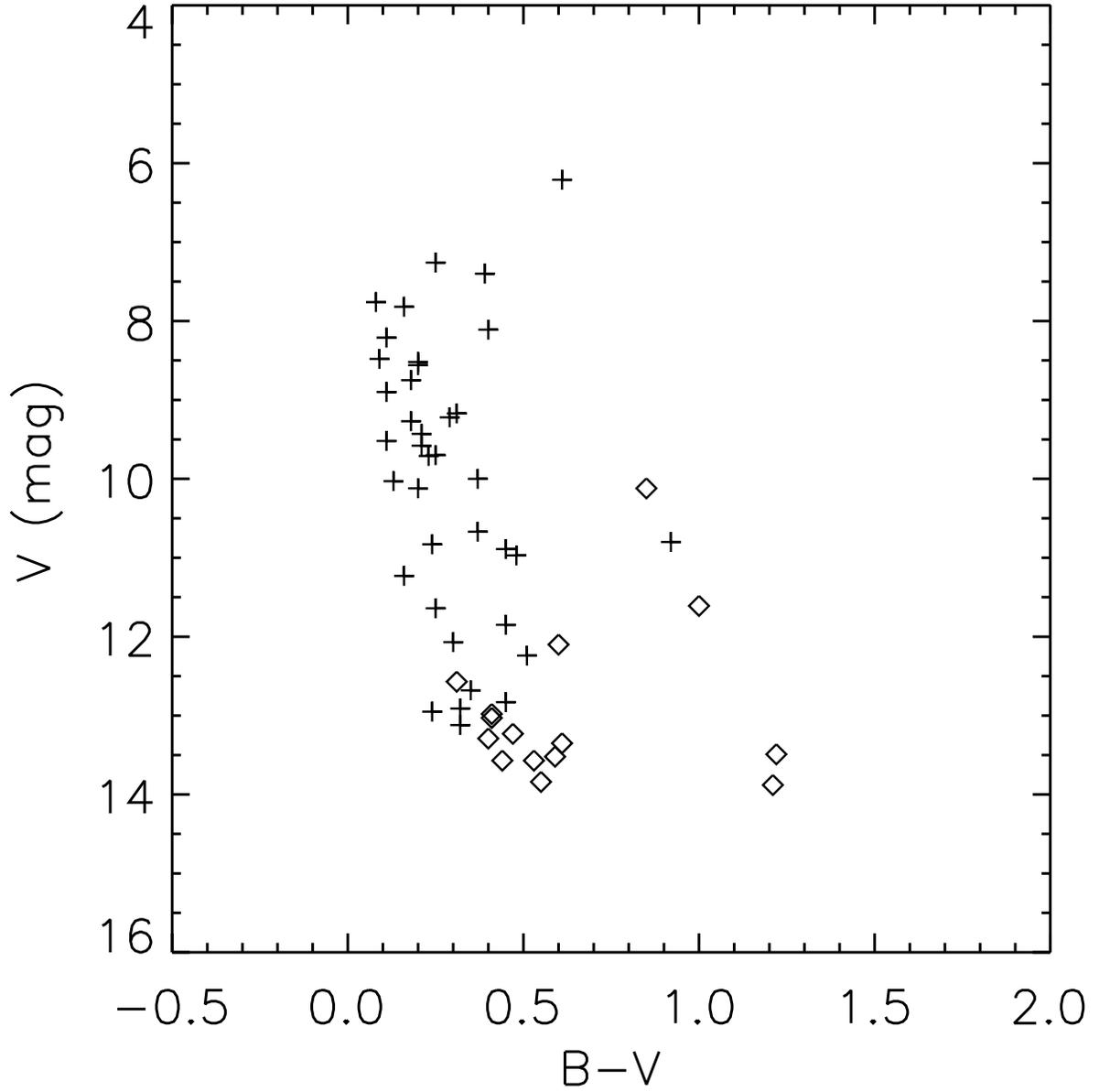}
\caption{
 The optical color magnitude diagram of the OB stars (pluses) and 
candidate OB stars (diamonds).}
\end{center}
\end{figure}

\clearpage

\begin{figure}
\begin{center}
\plotone{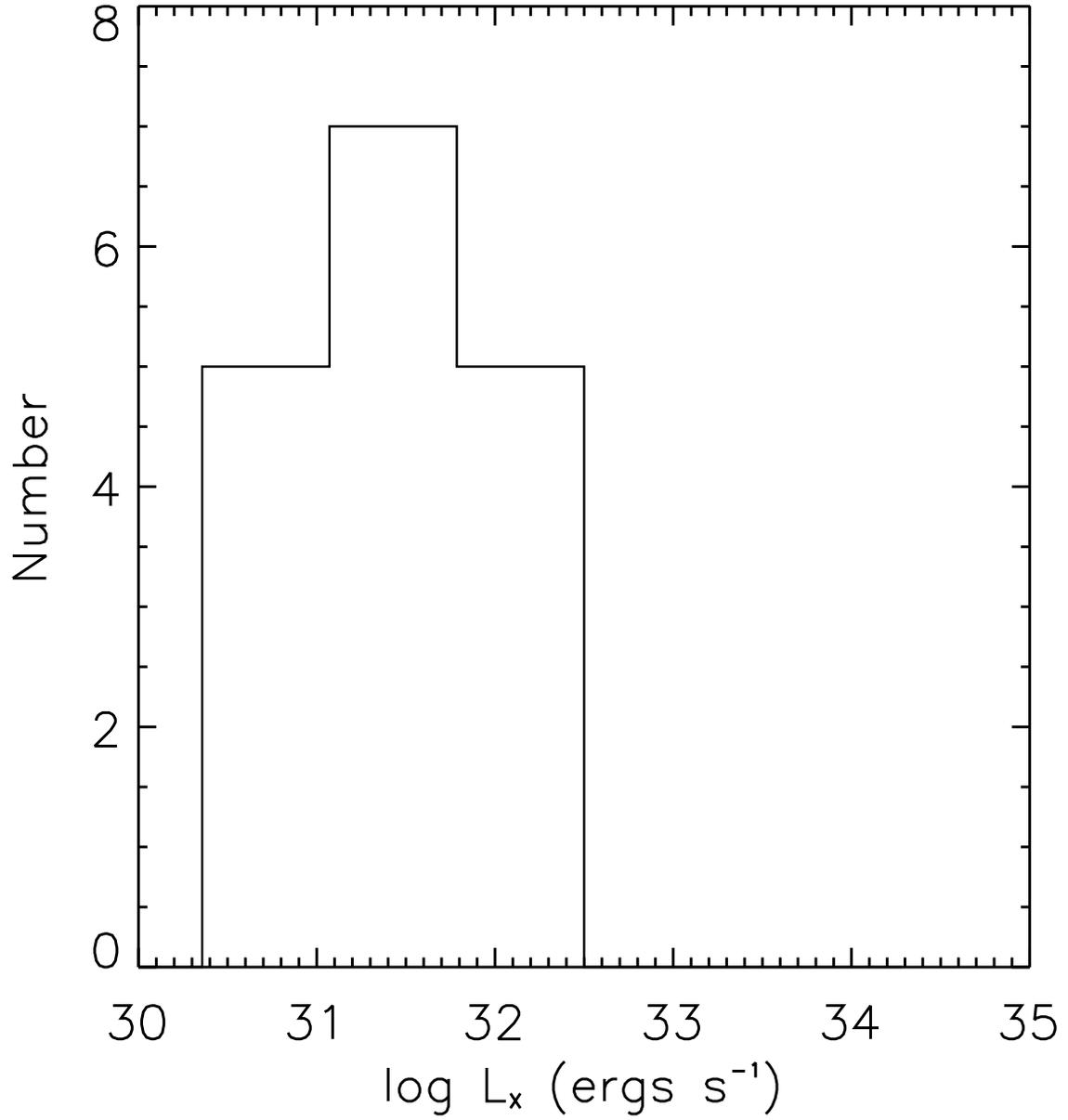}
\caption{
The distribution of X-ray luminosities of the candidate OB stars. 
The X-ray luminosities show similar range as for the 
known OB stars.}
\end{center}  
\end{figure}

\clearpage

\begin{figure}
\begin{center}
\plotone{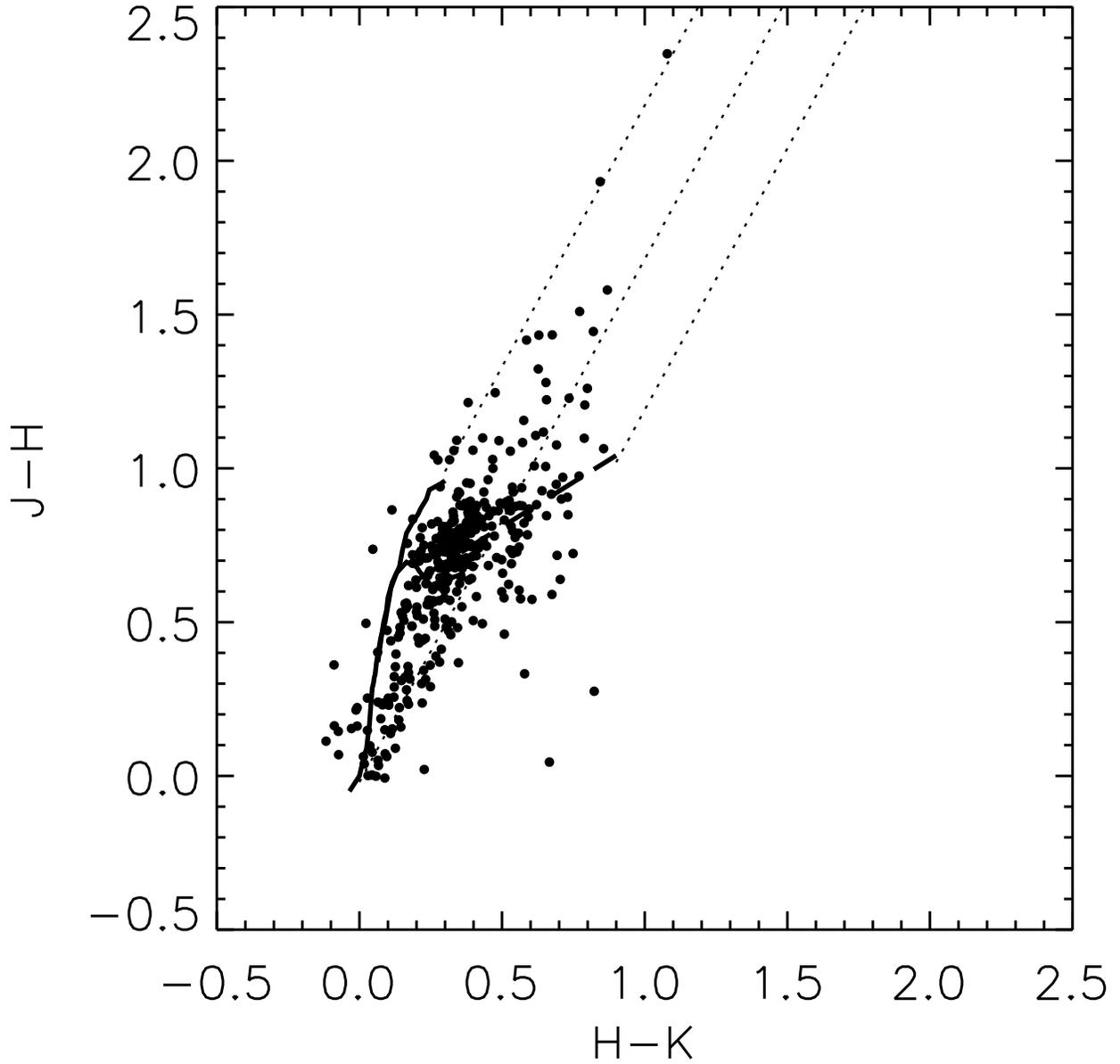}
\caption{The NIR color-color diagram of the X-ray sources. The dwarf and 
giant loci are shown as solid curves \citep{bb} and the classical 
T Tauri locus is shown as the dashed line \citep{meyer}. The dotted lines 
represent the reddening band.}
\end{center}
\end{figure}

\begin{figure}
\begin{center}
\plotone{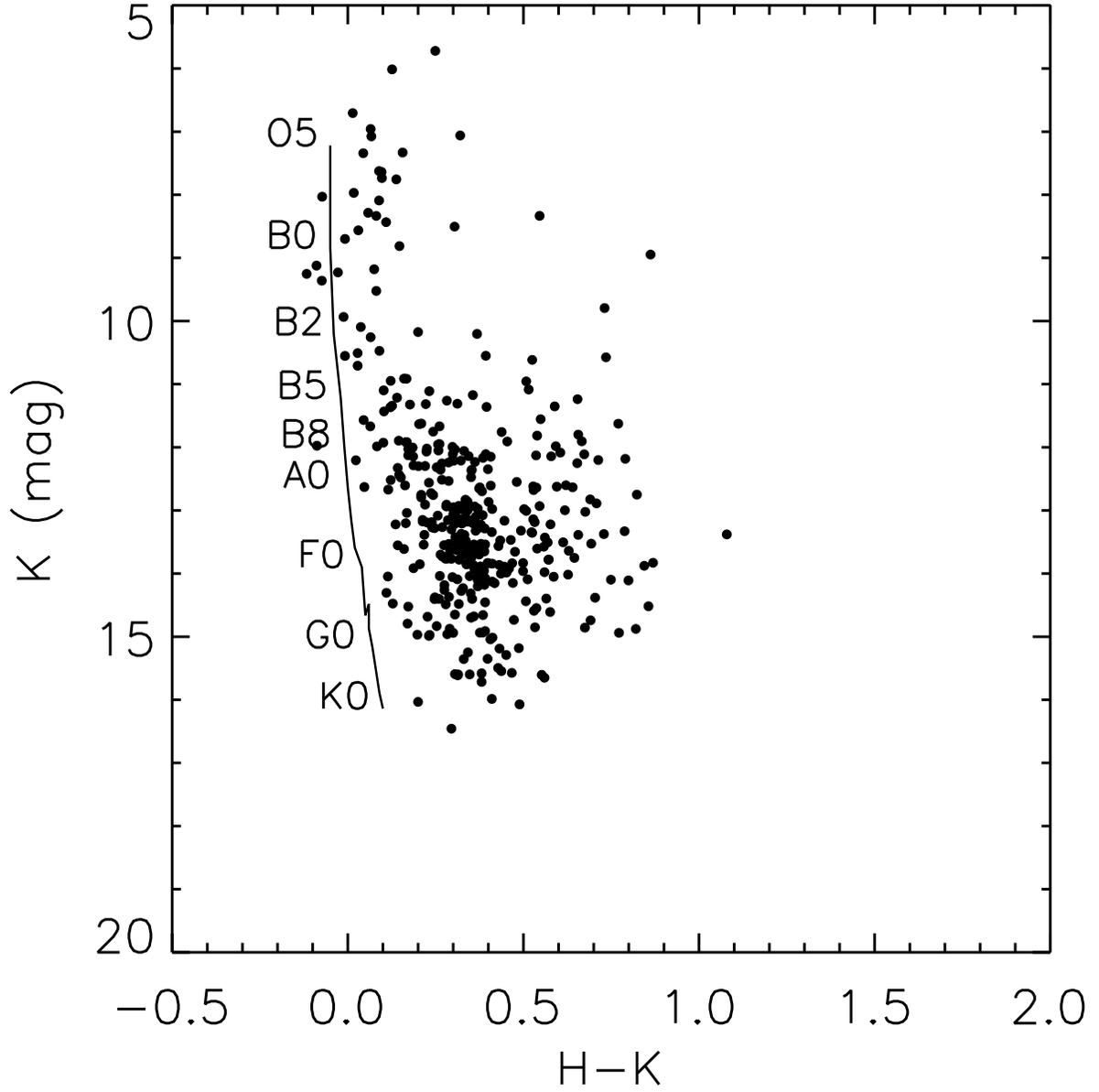}
\caption{
The NIR color-magnitude diagram of the X-ray sources. The vertical line 
represents the unreddened main sequence \citep{koornneef} at 2.5~kpc.}
\end{center}
\end{figure}

\begin{figure} 
\begin{center}
\plotone{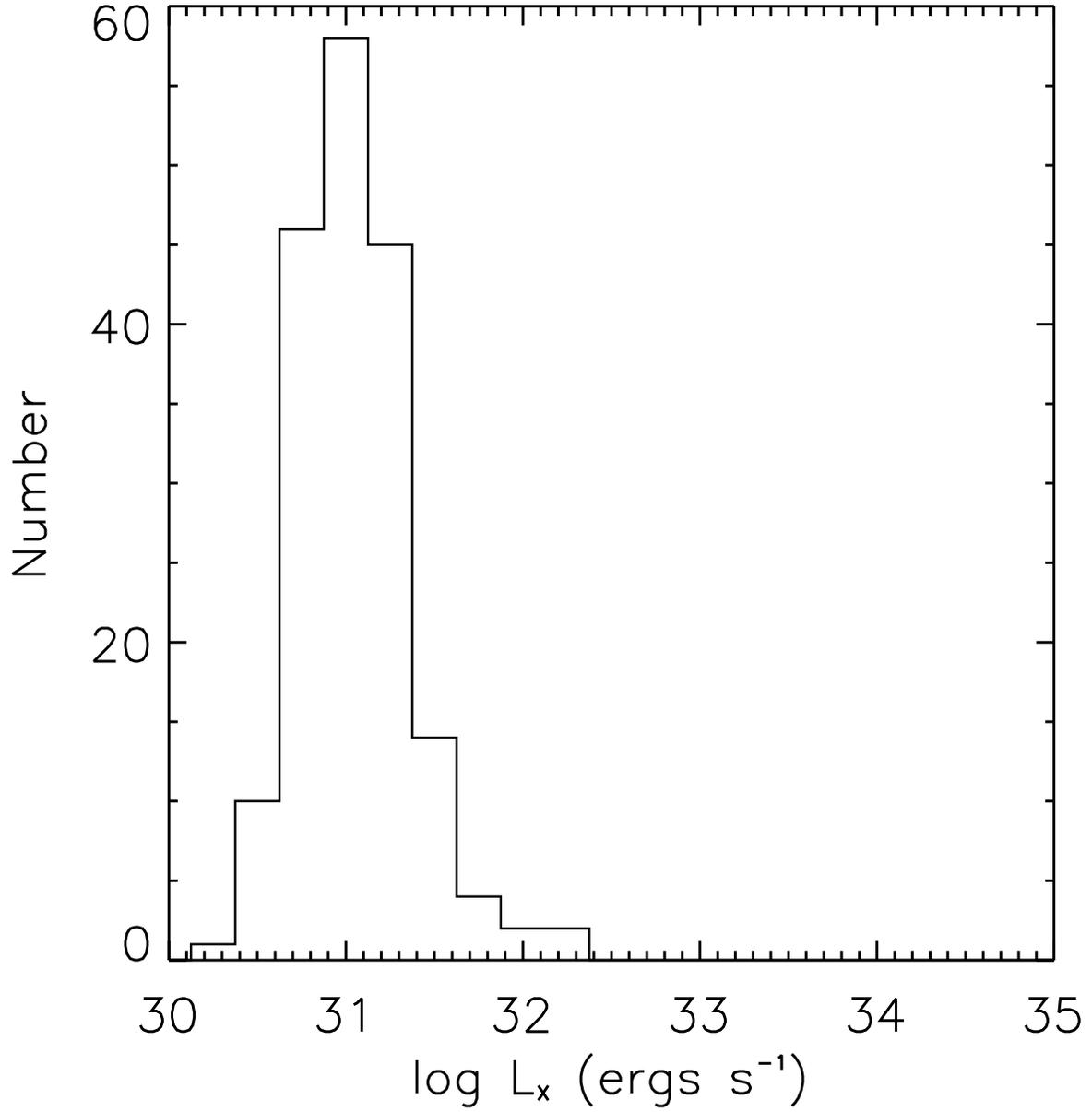} 
\caption{The distribution of X-ray luminosities of the T Tauri
candidates} 
\end{center} 
\end{figure}

\clearpage

\begin{figure}
\begin{center}
\plotone{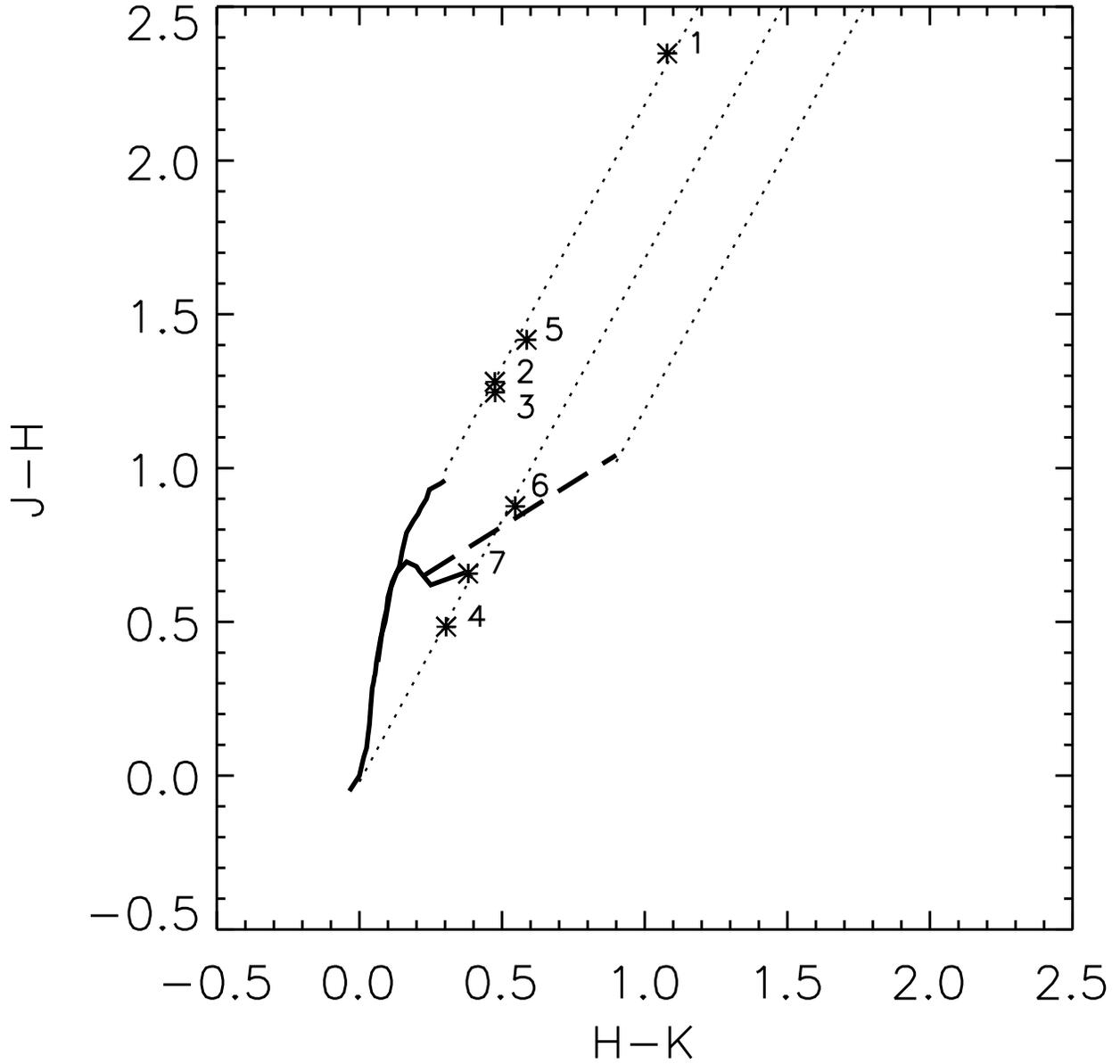}
\caption{The NIR color-color diagram of the embedded group of X-ray stars. 
Each star is labeled by its identification number in Table~6. The dwarf and
giant loci are shown as solid curves \citep{bb} and the classical
T Tauri locus is shown as the dashed line \citep{meyer}. The dotted lines
represent the reddening band.}
\end{center}
\end{figure}

\begin{figure}
\begin{center}
\plotone{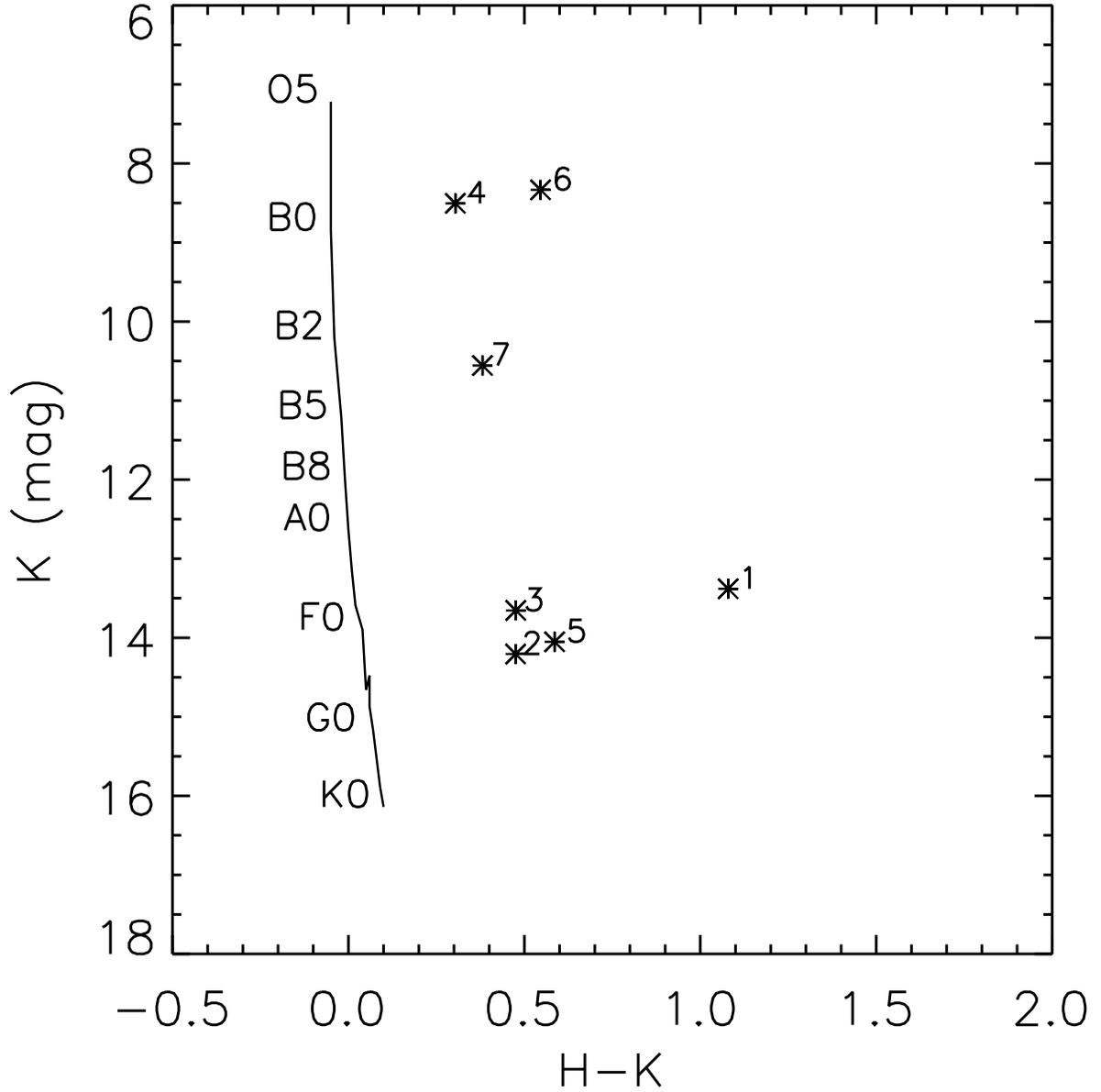}
\caption{The NIR color-magnitude diagram of the 
embedded group of X-ray stars.  Each star is labeled by 
its identification number in Table~6. The unreddened main-sequence track is 
plotted at 2.5 kpc \citep{koornneef}.} 
\end{center}
\end{figure}

\clearpage

\begin{figure}
\begin{center}
\plotone{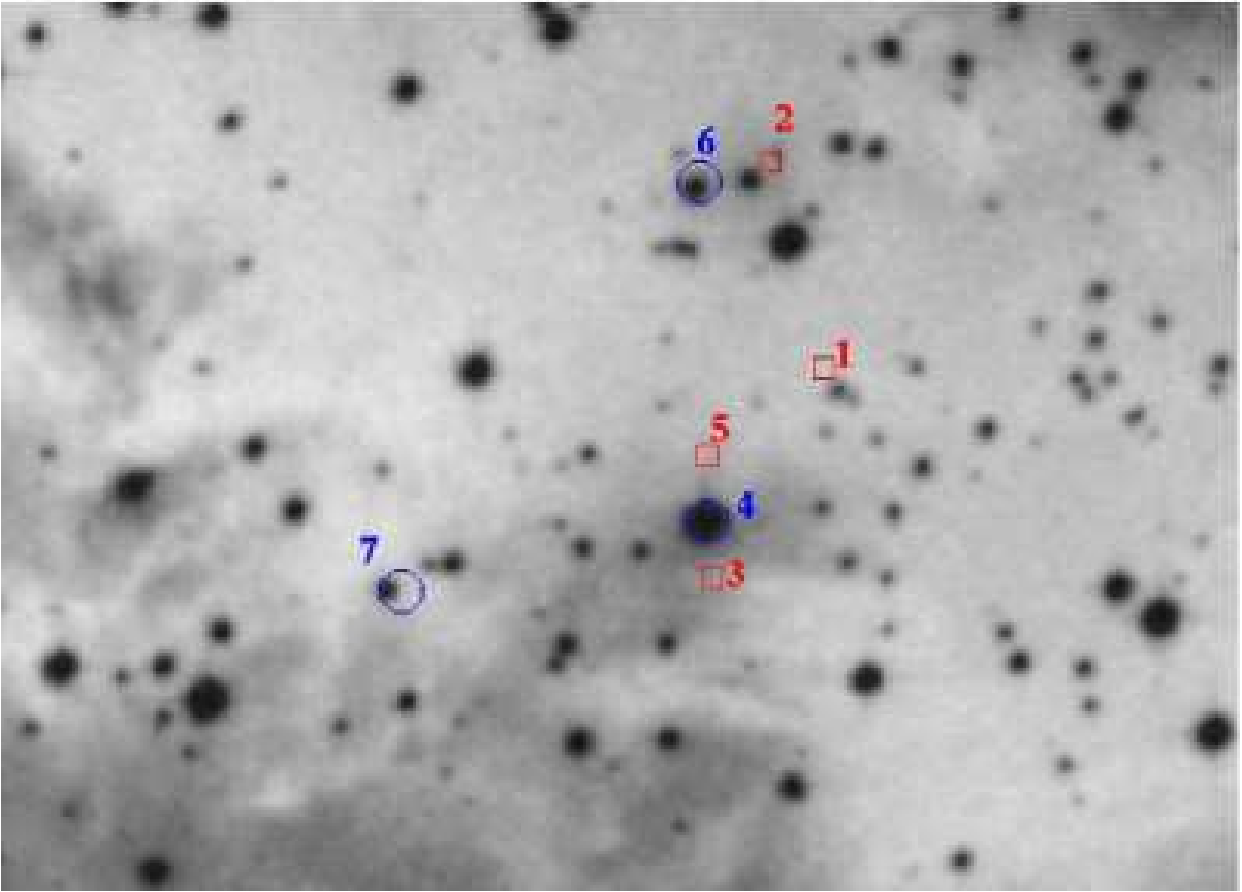}
\caption{The DSS image centered on, RA=$10^{h}45^{m}39.3^{s}$ , and 
DEC=$-59\arcdeg 48\arcmin 06\arcsec$, with a field of view of 
$5\arcmin \times 5\arcmin$.  The position of each X-ray source is  
marked with its ID number in Table~6. Only the 
bright NIR sources, 4, 6 and 7 have obvious counterparts.}    
\end{center}
\label{dss}
\end{figure}

\clearpage

\begin{figure}
\begin{center}
\plotone{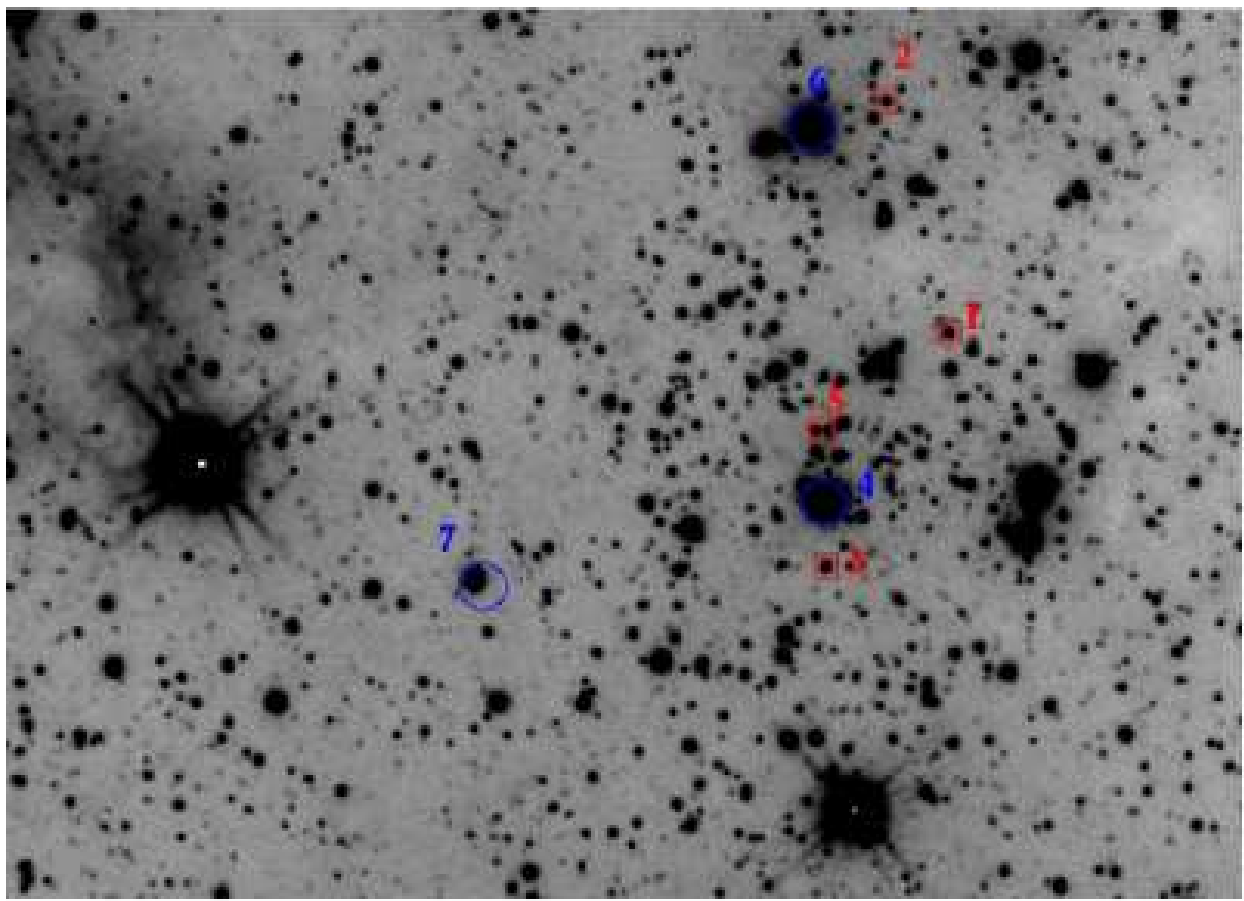}
\caption{
The SIRIUS $K_s$ image for the same field of view as Fig.~\ref{dss}.}
\end{center}
\end{figure}

\clearpage

\begin{figure}
\begin{center}
\plotone{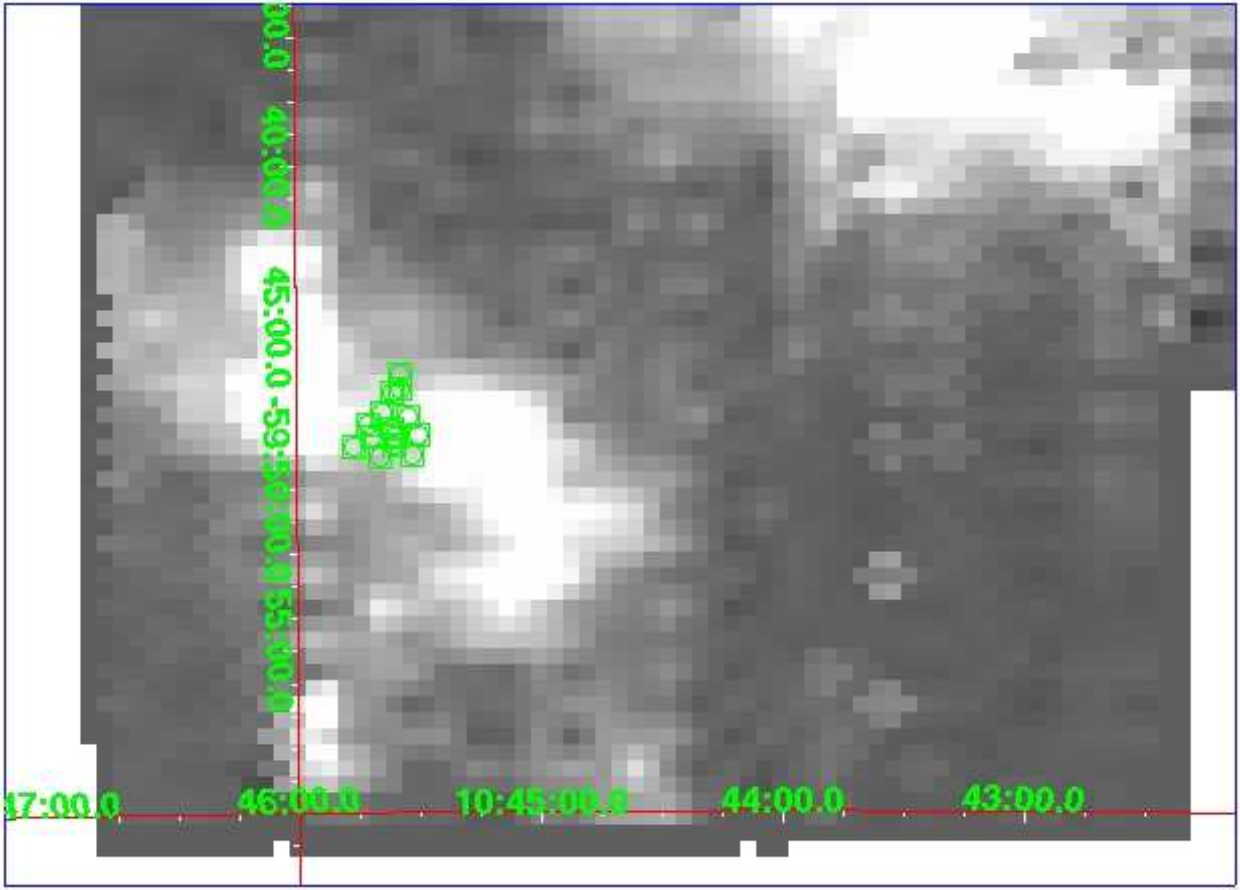}
\caption{
The X-ray group and the $~^{12}CO(1-0)$ emission \citep{brooks} in both 
contours and gray scale.} 
\end{center}  
\end{figure}

\begin{figure}
\begin{center}
\plotone{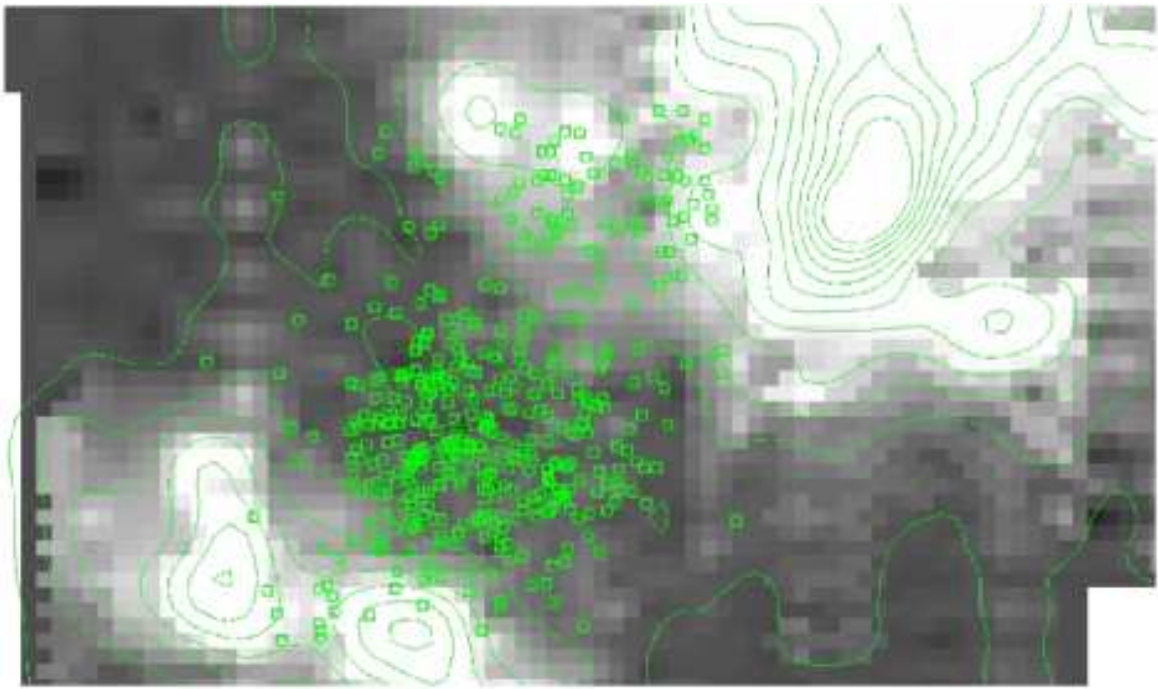}
\caption{The complete $Chandra$ sources and the $~^{12}CO(1-0)$ emission  
\citep{brooks} in both contours and gray scale.}
\end{center}  
\end{figure}

\end{document}

%% file: tab1.tex
\begin{deluxetable}{ccc}
\tablecaption{CHANDRA OBSERVATIONS}
\tablenum{1}
\tablewidth{0pt}
\tablehead{
\colhead{Observation ID} & \colhead{50} & \colhead{1249} \\
}
\startdata

Pointing              & 10$^{h}$45$^{m}$03$^{s}$.60 
-59$\arcdeg$41$\arcmin$03$\arcsec$ & 10$^{h}$45$^{m}$03$^{s}$.60 
-59$\arcdeg$41$\arcmin$03$\arcsec$ \\
Roll Observed         & 176$\arcdeg$        & 176$\arcdeg$         \\
Start Date            & 1999-09-06 19:48:12 & 1999-09-06 23:45:34  \\
Exposure Time         & 12.04~ks            & 9.75~ks              \\
Reduced Exposure Time & 8545.99~s           & 9574.01~s            \\
\enddata
\end{deluxetable}

%% file: tab2.tex
\begin{deluxetable}{cccrrccrrrrrccc}
\rotate
\tablecaption{Optical, NIR and X-ray data of known OB stars}
\tabletypesize{\scriptsize}
\tablenum{2}
\tablewidth{0pt}
\tablehead{
\colhead{R.A.} & \colhead{Decl.} & \colhead{Identifier} & \colhead{$B$} & \colhead{$V$} & \colhead{Sp. type}
& \colhead{$E(B-V)$} & \colhead{$M_{\rm {bol}}$} & \colhead{$J$} & \colhead{$H$} & \colhead{$K_s$} &
\colhead{$Chandra$} & \colhead{log Flux} & \colhead{log $L_{\rm X}$} 
& \colhead{log $L_{\rm X}/L_{\rm {bol}}$} \\ 

\colhead{h m s} & \colhead{\arcdeg~ \arcmin~ \arcsec} & \colhead{$ ~$} &
\colhead{(mag)} & \colhead{(mag)} & \colhead{$ ~$} & \colhead{$ ~$}  & \colhead{$
 ~$} & \colhead{(mag)} & \colhead{(mag)} & \colhead{(mag)} & \colhead{counts} &
\colhead{(ergs $\mathrm{~cm^{-2}~s^{-1}}$)} & \colhead{(ergs $\mathrm{s^{-1}}$)} &
\colhead{$~ $} \\

\colhead{(1)} & \colhead{(2)} & \colhead{(3)} & \colhead{(4)} & \colhead{(5)} &
\colhead{(6)} & \colhead{(7)} & \colhead{(8)} & \colhead{(9)} & \colhead{(10)} &
\colhead{(11)} & \colhead{(12)} & \colhead{(13)} & \colhead{(14)} & \colhead{(15)}
}
\startdata
 10 43 56.80 & -59 34 17.4 & Tr 14-25 & 13.44 & 13.12 & B5  & 0.49 & -2.6 & 11.94 & 11.96 & 11.76 & 71.5 & -13.06 & 31.83 & -4.76 \\
 10 43 57.51 & -59 32 53.1 & HD 93129 A/B & 7.51 & 7.26 & O3Iab & 0.55 & -11.5 & 6.23 & 6.14 & 6.01 & 2606.2 & -11.45 & 33.44 & -6.62 \\
 10 43 57.68 & -59 33 53.7 & Tr 14-18 & 12.37 & 12.07 & B0V & 0.60 & -6.9 & 10.76 & 10.54 & 10.55 & 22.7 & -13.47 & 31.42 & -6.81 \\
 10 43 59.96 & -59 32 26.3 & CD 58 3529 & 9.45 &  9.27 & O7V & 0.52 & -8.1 & 8.68 & 8.62 &  8.57 & 350.6 & -12.34 & 32.55 & -6.16 \\
 10 44 00.97 & -59 35 46.2 & Tr 16-200 & 11.04 & 10.67 & O9V & 0.48 & -7.6 & 9.44 & 9.25 &  9.18 & 39.3 & -13.32 & 31.56 & -6.94 \\
 10 44 05.04 & -59 33 42.6 & Tr 14-218 & 12.30 & 11.85 & B1.5V & 0.69 & -4.9 & 10.11 & 9.74  & 9.63 & 26.4 & -13.34 & 31.55 & -5.87 \\
 10 44 07.29 & -59 34 31.1 & Tr 14 229 & 7.98 & 7.82 & O6III & 0.31 & -9.3 & 7.17 & 7.14 & 7.08 & 436.8 & -12.42 & 32.47 & -6.71 \\
 10 44 08.99 & -59 34 34.0 & HD 93160 & 8.76 & 8.56 & O6.5V & 0.31 & -7.7 & 7.07 & 7.02 &  6.96 & 557.0 & -12.31 & 32.58 & -5.97 \\
 10 44 10.44 & -59 43 11.2 & HD 93162 & 8.51 & 8.10 & WN+ & 0.66 & -10.1 & 6.26 & 5.97 & 5.72 & 10179.0 & -10.77 & 34.12 & -5.39  \\
 10 44 13.26 & -59 43 10.4 & Tr 16-244 & 11.72 & 10.80 & O3.5Iab & 0.94 & -8.6 & 7.84 &  7.38 &  7.06 & 226.4 & -12.23 & 32.66 & -6.25 \\
 10 44 22.55 & -59 39 25.9 & Tr 16-11 & 11.39 & 11.23 & B1.5 & 0.43 & -4.9 & 10.63 & 10.56 & 10.47 & 16.5 & -13.74 & 31.14 & -6.28 \\
 10 44 30.44 & -59 37 26.8 & Tr 16-10 & 9.94 &  9.71 & B0Vn & 0.57 & -6.9 & 8.84 & 8.67 & 8.57 &  43.8 & -13.21 & 31.68 & -6.54 \\
 10 44 32.39 & -59 44 31.2 & HD 93204 &8.57 &  8.40 & O5V((f)) & 0.41 & -9.0 & 8.02 & 7.98 &  7.97 & 192.3 & -12.69 & 32.30 & -6.87 \\
 10 44 33.80 & -59 44 15.6 & HD 93205 & 7.84 & 7.76 & O3V & 0.40 & -10.1 & 7.38 & 7.38 & 7.34 & 985.5 & -11.99 & 32.90 & -6.60 \\
 10 44 36.78 & -59 47 29.7 & CPD -59$^{o}$ 2591 & 11.34 & 10.89 & O8V & 0.76 & -7.1 & 9.39 & 9.03 &  9.12 & 15.9 & -13.50 & 31.38 & -6.92 \\
 10 44 37.43 & -59 32 56.0 & HDE 303311 & 9.01 &  8.90 & O5V & 0.45 & -8.6 & 8.59 & 8.59 &  8.56 & 124.1 & -12.85 & 32.04 & -6.86 \\
 10 44 41.87 & -59 46 56.5 & CPD -59$^{o}$ 2600 & 8.72 &  8.50 & O6V((f)) & 0.51 & -9.0 & 7.79 &  7.73 &  7.64 &  289.2 & -12.43 & 32.46 & -6.61 \\
 10 44 44.22 & -59 42 33.6 & Tr 16-385 & 13.23 & 12.91 & B5 & 0.49 & -2.6 & 11.76 & 11.53 & 11.43 & 35.6 & -13.36 & 31.53 & -4.97 \\
 10 44 45.06 & -59 33 55.2 & HD 93250 & 7.79 & 7.40 & O3V & 0.49 & -10.7 & 6.78 & 6.72 & 6.71 & 2927.9 & -11.44 & 33.43 & -6.30 \\
 10 44 46.03 & -59 40 32.3 & Tr 16-395 & 12.75 & 12.24 & B7 & 0.62 & -1.0 & 10.99 & 10.73 & 10.70  & 8.0 & -13.91 & 30.98 & -4.89 \\
 10 44 47.36 & -59 43 53.4 & CPD -59$^{o}$ 2603 & 8.93 & 8.75 & O7V((f)) & 0.46 & -8.5 & 8.34 &  8.34 &  8.29 & 108.0 & -12.90 & 31.99 & -6.88 \\
 10 44 54.12 & -59 41 29.6 & CPD -59$^{o}$ 2606 & 11.07 & 10.83 & B2:Vn & 0.43 & -4.8 & 10.23 & 10.13 & 10.09 &  9.0 & -14.01 & 30.88 & -6.50 \\
 10 44 57.83 & -59 39 59.7 & Tr 16-453 & 13.28 & 12.83 & B3 & 0.64 & -4.8 & 11.32 & 11.08 & 10.91 & 11.7 & -13.72 & 31.16 & -6.22 \\
 10 45 00.06 & -59 43 34.6 & CPD -59$^{o}$ 2616 & 11.89 & 11.64 & B3 & 0.44 & -4.8 & 10.68 & 10.53 & 10.50  & 20.7 & -13.63 & 31.26 & -6.13 \\
 10 45 03.54 & -59 41 04.2 & $\eta$ Carinae & 6.2 & 5.4 & LBV & 0.40 & -12.0 & 1.65 & 1.07 & 0.17 & 808.9 & -12.07 & 32.76 & -7.54 \\
 10 45 05.86 & -59 45 19.7 & Tr 16-23 & 10.37 & 10.00 & O7V & 0.69 & -7.5 & 8.64 & 8.41 &  8.34 & 45.8 & -13.10 & 31.79 & -6.67 \\
 10 45 05.88 & -59 43 07.8 & Tr 16-9 & 9.95 &  9.70 & O9.5V & 0.56 & -7.2 & 8.85 & 8.69 &  8.70 &  55.0 & -13.11 & 31.77 & -6.57 \\
 10 45 05.95 & -59 40 06.3 & HDE 303308 & 8.32 & 8.21 & O3V((f)) & 0.46 & -9.8 & 7.70 & 7.71 & 7.62 & 440.4 & -12.29 & 32.60 & -6.79 \\
 10 45 06.76 & -59 41 56.8 & Tr 16-3 & 10.32 & 10.12 & O8.5V & 0.52 & -6.9 & 9.43 & 9.28 &  9.36 & 36.0 & -13.33 & 31.56 & -6.67 \\
  10 45 08.26 & -59 40 49.7 & CPD -59$^{o}$ 2628 & 9.63 &  9.52 & O9.5V & 0.44 & -7.5 & 9.24 & 9.13 &   9.25 & 67.8 & -13.12 & 31.77 & -6.70 \\
 10 45 08.26 & -59 46 07.3 & Tr 16-22 & 11.45 & 10.97 & O8.5V & 0.78 & -6.8 & 9.27 & 8.96 &  8.81 & 463.2 & -12.03 & 32.86 & -5.32 \\
 10 45 08.31 & -59 38 47.5 & Tr 16-492 & 13.03 & 12.68 & B3 & 0.52 & -4.8 & 11.39 & 11.07 & 10.94 & 20.0 & -13.59 & 31.30 & -6.09 \\
 10 45 08.75 & -59 40 42.3 & Tr 16-495 & 13.19 & 12.95 & B5 & 0.41 & -2.6 & 12.05  & 11.88 & 11.97 & 7.0 & -14.13 & 30.76 & -5.74 \\
 10 45 12.28 & -59 45 00.6 & HD 93343 & 9.79 &  9.58 & O8V & 0.56 & -8.0 & 8.68 & 8.54 &  8.43 & 117.4 & -12.79 & 32.10 & -6.56 \\
 10 45 12.78 & -59 44 46.4 & CPD -59$^{o}$ 2635 & 9.64 &  9.43 & O8.5V & 0.54 & -8.0 & 8.33 & 8.18 &   8.09 & 98.6 & -12.88 & 32.01 & -6.66 \\
 10 45 12.93 & -59 44 19.4 & CPD -59$^{o}$ 2636 & 9.48 &  9.17 & O8.5V & 0.60 & -8.3 & 8.07 & 7.89 &  7.76 & 191.0 & -12.56 & 32.33 & -6.46 \\
 10 45 16.58 & -59 43 37.3 & CPD -59$^{o}$ 2641 & 9.51 &  9.22 & O6V((f)) & 0.61 & -8.7 & 8.02 &   7.95 &  8.03  & 159.7 & -12.61 & 32.27 & -6.67 \\
 10 45 20.60 & -59 42 51.5 & Tr 16-115 & 10.16 & 10.03 & O8.5V & 0.46 & -6.8 & 9.35 & 9.20 &  9.23 & 30.7 & -13.45 & 31.44 & -6.74 \\
\enddata
\end{deluxetable}

%% file: tab3.tex
\begin{deluxetable}{cccrrrrrrcc}
\tablecaption{Optical, NIR and X-ray data of candidate OB stars}
\rotate
\tablenum{3}
\tablewidth{0pt}
\tablehead{
\colhead{ID} & \colhead{R.A.} & \colhead{DEC.} & \colhead{$B$} & \colhead{$V$} & \colhead{$J$}
& \colhead{$H$} & \colhead{$K_s$} & \colhead{$Chandra$} &
\colhead{log Flux} & \colhead{log $L_{\rm X}$} \\ 

\colhead{$ ~$} & \colhead{h m s} & \colhead{\arcdeg~ \arcmin~ \arcsec} &
\colhead{(mag)} & \colhead{(mag)} & \colhead{(mag)} & \colhead{(mag)} & 
\colhead{(mag)} & \colhead{counts} & \colhead{(ergs $\mathrm{~cm^{-2}~s^{-1}}$)} & 
\colhead{(ergs $\mathrm{s^{-1}}$)} \\

\colhead{(1)} & \colhead{(2)} & \colhead{(3)} & \colhead{(4)} & \colhead{(5)}
& \colhead{(6)} & \colhead{(7)} & \colhead{(8)} & \colhead{(9)} & \colhead{(10)}
& \colhead{(11)} 
}
\startdata
1 & 10 44 02.71 & -59 39 45.6 & 14.71 &  13.49 & 10.90 & 10.37 & 10.17 & 98.4 & -12.94 & 31.90 \\
2 & 10 44 06.74 & -59 36 10.9 & 13.69 &  13.29 & 11.65 & 11.34 & 11.11 & 118.5 & -12.18 & 32.70 \\
3 & 10 44 27.73 & -59 45 21.3 & 10.97 &  10.12 & 8.30 & 7.83 & 7.73 & 70.8 &   -12.57 &  32.31 \\
4 & 10 44 29.73 & -59 32 20.2 & 12.61 &  11.61 & 7.99 & 7.48 & 7.33 &  64.6 &  -12.52 & 32.35  \\
5 & 10 44 33.61 & -59 38 20.7 & 14.10 &  13.57 & 11.82 & 11.46 & 11.34 & 7.7 & -14.04 & 30.79 \\
6 & 10 44 46.53 & -59 34 13.6 & 13.70 &  13.23 & 11.82 & 11.40 & 11.30 & 146.5 &  -12.63  & 32.26 \\
7 & 10 44 57.01 & -59 38 26.6 & 15.09 &  13.88 & 11.63 & 11.07 & 10.91 & 46.5 &  -12.64 & 32.25 \\
8 & 10 44 58.35 & -59 39 43.6 & 13.39 &  12.98 & 11.74 & 11.48 & 11.36 & 5.0 &   -14.14 &  30.74  \\
9 & 10 45 07.84 & -59 41 34.0 & 13.44 &  13.03 & 11.57 & 11.35 & 11.21 & 27.0 &  -13.41 & 31.48 \\
10 & 10 45 11.16 & -59 42 33.5 & 14.39 &  13.84 & 11.92 & 11.46 & 10.95 & 19.0 & -13.66 & 31.18 \\
11 & 10 45 16.18 & -59 41 41.1 & 14.01 &  13.57 & 11.97 & 11.53 & 11.31 & 39.0 &  -13.26 & 31.65 \\
12 & 10 45 18.81 & -59 42 17.9 & 12.88 &  12.57 & 11.43 & 11.20 & 11.09 & 17.0  & -13.71 & 31.18 \\
13 & 10 45 21.81 & -59 45 25.0 & 13.96 &  13.35 & 11.81 & 11.50 & 11.32 & 11.4 &  -13.64 & 31.25 \\
14 & 10 45 36.40 & -59 48 24.2 & 12.70 &  12.10 & 9.29 & 8.80 & 8.50 & 75.8 &  -12.77 & 32.12 \\
15 & 10 45 36.79 & -59 47 02.0 & \nodata & \nodata & 9.75 & 8.88 &  8.33 & 47.1 & -13.26 & 31.57 \\
16 & 10 45 38.28 & -59 42 07.7  & 14.11 &  13.52 & 11.32 & 10.57 & 10.20 & 29.5 &  -13.24 & 31.64 \\
17 & 10 45 46.36 & -59 48 42.1 &  \nodata & \nodata & 11.59  & 10.93 & 10.55 & 55.4 & -13.19 & 31.64 \\
\enddata
\end{deluxetable}

%% file: tab4.tex
\begin{deluxetable}{cccrrrr}
\tablecaption{The NIR magnitudes and X-ray counts of the X-ray group 
stars}
\tablenum{4}
\tablewidth{0pt}
\tablehead{
\colhead{ID} & \colhead{R.A.} & \colhead{DEC.} & \colhead{$J$} & 
\colhead{$H$} & \colhead{$K_s$}
& \colhead{$Chandra$} \\

\colhead{$ ~$} & \colhead{h m s} & \colhead{\arcdeg~ \arcmin ~\arcsec} &
\colhead{(mag)} & \colhead{(mag)} & \colhead{(mag)} & \colhead{(counts)} \\

 }
\startdata
1 & 10 45 32.73 & -59 47 47.0 & 16.80 & 14.45 & 13.38 & 28.7 \\
2 & 10 45 34.52 & -59 46 56.3 & 15.96 & 14.68 & 14.20 & 6.1  \\
3 & 10 45 36.32 & -59 48 38.0 & 15.37 & 14.13 & 13.65 & 3.4  \\
4 & 10 45 36.40 & -59 48 24.2 & 9.29 & 8.80 & 8.50 & 75.8  \\
5 & 10 45 36.46 & -59 48 08.3 & 16.05 & 14.63 & 14.05 & 82.9  \\
6 & 10 45 36.79 & -59 47 02.0 & 9.75 & 8.88 & 8.33 & 47.1  \\
7 & 10 45 46.36 & -59 48 42.1 & 11.59 & 10.93 & 10.55 & 55.4 \\
\enddata
\end{deluxetable}